\documentclass[draftclsnofoot,onecolumn]{IEEEtran}

\usepackage{cite}
\usepackage{color}

\ifCLASSINFOpdf
  \usepackage[pdftex]{graphicx}
  \graphicspath{{../pdf/}{../jpeg/}}
\else
\fi

\usepackage{amsmath}
\usepackage{amsfonts}
\usepackage{cases}
\usepackage{amssymb}
\usepackage{bm}
\usepackage[justification=centering]{caption}
\usepackage{titlecaps}

\usepackage{multirow}
\usepackage{threeparttable}
\usepackage{enumerate}
\usepackage{subfigure}
\makeatother
\usepackage[export]{adjustbox} 
\usepackage{booktabs}
\usepackage{setspace}
\usepackage{xcolor}
\usepackage{mathrsfs}
\usepackage[ruled,linesnumbered]{algorithm2e}
\makeatletter
\newcommand{\nosemic}{\renewcommand{\@endalgocfline}{\relax}}
\newcommand{\dosemic}{\renewcommand{\@endalgocfline}{\algocf@endline}}
\let\oldnl\nl
\newcommand{\nonl}{\renewcommand{\nl}{\let\nl\oldnl}}
\makeatother
\usepackage[export]{adjustbox} 
\usepackage{booktabs}
\usepackage{setspace}
\allowdisplaybreaks[3]
\usepackage{caption}
\usepackage{floatrow}
\usepackage{etoolbox}

\usepackage{pgfplots}
\pgfplotsset{compat=newest}

\usepackage{url}

\usepackage{xcolor}

\allowdisplaybreaks[4]

\usepackage{comment}
\usepackage{ifthen}

\begin{document}

\title{\LARGE A Distributed Incremental Update Scheme for Probability Distribution of Wind Power Forecast Error}
\author{Mengshuo Jia$^{\dagger, \star}$, Chen Shen$^{\dagger, \star, \diamond}$, Zhaojian Wang$^{\dagger, \star}$}
\maketitle

\vspace{-1.5cm}
\centerline{$^{\dagger}$\textit{\small{Department of Electrical Engineering, Tsinghua University,  Beijing 100084, China}}}

\centerline{$^{\star}$\textit{\small{State Key Laboratory of Power Systems, Beijing 100084, China}}}

\centerline{$^{\diamond}$\textit{\small{Author for Correspondence: shenchen@mail.tsinghua.edu.cn}}}

\vspace{10pt}

\noindent \textbf{Abstract:} 
Due to the uncertainty of distributed wind generations (DWGs), a better understanding of the probability distributions (PD) of their wind power forecast errors (WPFEs) can help market participants (MPs) who own DWGs perform better during trading. Under the premise of an accurate PD model, considering the correlation among DWGs and absorbing the new information carried by the latest data are two ways to maintain an accurate PD. These two ways both require the historical and latest wind power and forecast data of all DWGs. Each MP, however, only has access to the data of its own DWGs and may refuse to share these data with MPs belonging to other stakeholders. Besides, because of the endless generation of new data, the PD updating burden increases sharply. Therefore, we use the distributed strategy to deal with the data collection problem. In addition, we further apply the incremental learning strategy to reduce the updating burden. Finally, we propose a distributed incremental update scheme to make each MP continually acquire the latest conditional PD of its DWGs' WPFE. Specifically, we first use the Gaussian-mixture-model-based (GMM-based) joint PD to characterize the correlation among DWGs. Then, we propose a distributed modified incremental GMM algorithm to enable MPs to update the parameters of the joint PD in a distributed and incremental manner. After that, we further propose a distributed derivation algorithm to make MPs derive their conditional PD of WPFE from the joint one in a distributed way. Combining the two original algorithms, we finally achieve the complete distributed incremental update scheme, by which each MP can continually obtain its latest conditional PD of its DWGs' WPFE via neighborhood communication and local calculation with its own data. The effectiveness, correctness, and efficiency of the proposed scheme are verified using the dataset from the NREL.

\vspace{10pt}
\vspace{10pt}


\vspace{10pt}
\noindent \textbf{1. Introduction} 
\vspace{5pt}

Significantly increasing development of distributed renewable energy (DRE) allows traditionally passive consumers to become `prosumers' under consumer level communications and control \cite{morstyn2018using}. Accordingly, different trading schemes have emerged to incentivise prosumers to participate in local markets, e.g., peer-to-peer scheme \cite{LUTH20181233} and aggregator scheme \cite{LIU2018689}. No matter which scheme market participants (MPs) choose, they have to face the uncertainty brought by their DREs \cite{7930435}. Since generating a perfect forecast remains a challenge under current technologies \cite{6481495}, a better understanding of the probability distribution (PD) of the DRE forecast error can help these MPs perform better in storage system arrangement \cite{WU2014100} and bidding strategy selection \cite{1490597}. In this paper, we focus on distributed wind generation (DWG) and aim to make each MP obtain an accurate PD of its DWGs' forecast error.

There are two ideas that can help MPs build and maintain an accurate PD of wind power forecast error (WPFE). One is considering the wind power correlation among DWGs when building the PD. Another one is updating the established PD in real time by absorbing new information carried by new data. It should be emphasized that, knowing the wind power forecast values of all DWGs may assist each MP to build a more accurate distribution than the one it individually creates with its own data \cite{EXIZIDIS201665}. Therefore, each MP should first build the joint PD of all DWGs' wind power and forecast data. Then, each MP should derive the conditional PD of its DWGs' WPFE under all DWGs' latest forecast values from the joint one \cite{7434078,7862254,wang2018}. Meanwhile, each MP should update its conditional PD when new wind power and forecast data emerge. The prerequisite of the two ideas is that each MP can always have access to the historical and latest wind power and forecast data of all DWGs. The fact, however, is that each MP only has access to the data of its own DWGs. As the bidding of DWGs is dependent on their forecast values, MPs may refuse to share these data with MPs belonging to other stakeholders. To avoid the data collection, a distributed strategy is a feasible choice. Except for the problem of data collection, each MP also faces the high calculation burden when updating the conditional PD due to the endless increase in new data. To solve the two problems, this paper aims to develop an efficient distributed update scheme to make each MP obtain the latest conditional PD of its DWGs' WPFE. For the sake of brevity, we will hereinafter use this phrase, `each MP's conditional PD of WPFE', to represent each MP's conditional PD of its DWGs' WPFE when it does not cause ambiguity.


A suitable distribution model is the basis of the scheme. Many efforts have been made to model the PD of the WPFE. A common approach is to represent the WPFE as a normally distributed random variable with the forecast value as the expected value \cite{4470561}. The Weibull distribution, beta distribution \cite{1490597}, and Cauchy distribution \cite{6039388} are other common models for characterizing the uncertainty of the WPFE. However, the Gaussian, beta, and Cauchy distributions are not universal, particularly when the time scales change \cite{7779540}. Other modified models, e.g., the Levy alpha-stable distribution \cite{6822599}, mixed beta distribution \cite{5765544}, Gamma-like distribution \cite{6202398}, and t-location distribution \cite{6074302}, have also been proposed to better fit the features of the WPFE. Subsequently, to improve the generality of the PD model, researchers have tended to utilize or develop models that have more adjustable parameters. Zhang et al. presented a versatile distribution with three adjustable shape parameters estimated by nonlinear least-squares \cite{6481495,6344672}, which not only can better represent the WPFE than the Gaussian and beta distributions \cite{6481495} but also has analytical forms of the cumulative distribution function (CDF) and its inverse function. However, the unimodal feature of the versatile distribution limits its application \cite{7935491}. An improved versatile distribution with a more accurate representation was first proposed to solve this problem. Then, a versatile mixture distribution was developed through the convex combination of several improved versatile distributions \cite{7935491}. Recently, the Gaussian mixture model (GMM) has been applied to stochastic analyses of power systems because it can characterize multidimensional random variables subject to an arbitrary distribution with high precision regardless of the timescale \cite{5298967,8283770,7744686}. Meanwhile, the GMM makes it easy to derive the conditional PD from the GMM-based joint PD because of its highly analytical feature \cite{wang2018,8481551}. Thus, many researchers use GMM to represent the conditional PD of the WPFE \cite{7434078,7862254,wang2018}. The GMM is also adopted in this paper.  

The expectation--maximization (EM) algorithm is the most commonly used method to estimate the parameters of the GMM \cite{5298967,8283770}. This algorithm needs to collect the historical wind power and wind power forecast data of the correlated DWGs and train them. However, as an offline training method, this algorithm requires a complete historical dataset for each training. As new data continue to emerge, the size of the historical dataset increases, resulting in a sharply increasing update cost. To realize an efficient update scheme, the incremental learning algorithm is a feasible method. As a dynamic technique in machine learning, the incremental learning algorithm can learn new knowledge only from new data and keep most of the knowledge that has been learned before, thus avoiding the repeated training of old data and improving the efficiency. For updating the parameters in the GMM, incremental learning algorithms have been investigated in the fields of video recognition, robot control, etc. \cite{song2005highly,declercq2008online,chen2013incremental,bouchachia2011incremental,784637,6491172,pinto2015fast,pinto2017scalable}. The research on incremental learning algorithms can be divided into two types: (1) block-wise incremental learning algorithms \cite{song2005highly,declercq2008online,chen2013incremental} and (2) point-wise incremental learning algorithms  \cite{bouchachia2011incremental,784637,6491172,pinto2015fast,pinto2017scalable}. The former updates the GMM after a certain amount of new data has been generated, while the latter updates GMM point by point when a piece of new data is generated. The key to the block-wise incremental learning algorithm lies in whether and how to merge the two GMMs built by the old dataset and the new dataset. Merging judgments include the statistical feature equivalence judgment \cite{song2005highly}, fidelity judgment \cite{declercq2008online} and Kullback-Leibler divergence judgment \cite{chen2013incremental}, and the merging operations are based on weighted fusion. Since the block-wise incremental learning algorithm requires waiting for the generation of a certain amount of new data, the updating may not be timely. Therefore, the point-wise incremental learning algorithm is a better choice. The point-wise incremental learning algorithm generally consists of 3 steps: the judging step, updating step, and creating step. The first step is to decide whether a piece of new data belongs to an existing Gaussian component in the GMM. Different criteria are utilized: the likelihood value of a component when substituting the latest data in it \cite{6491172}, the distance between the latest data and the expected value of the components \cite{784637}, or the Mahalanobis distance between the latest data and the components \cite{bouchachia2011incremental,pinto2015fast,pinto2017scalable}. If the judgment passed, i.e., the new data belongs to an existing Gaussian component, then the parameters of the GMM will be updated based on the new data in the second step. The Robbins-Monro stochastic approximation method is the most common method to estimate the updated parameters of the GMM \cite{6491172} and is improved in \cite{pinto2015fast,pinto2017scalable} by introducing the inverse of the covariance to avoid the need for the inverse calculation in each update. The filter-based algorithms, i.e., the causal low-pass filter \cite{784637} and self-adapted recursive filter \cite{bouchachia2011incremental}, are also used to update the parameters. If the judgment is not passed, then a new Gaussian component is created \cite{6491172,pinto2015fast,pinto2017scalable}, or the less contributing component is replaced \cite{bouchachia2011incremental} in the third step with a preset covariance and a mean vector equal to the new data.

Among the above point-wise incremental learning algorithms, we choose the commonly used modified incremental GMM (IGMM) algorithm \cite{pinto2015fast} to realize an incremental update scheme for the GMM-based PD. This scheme, however, still needs to collect the continuously new wind power and forecast data from all DWGs. To deal with the data barriers among MPs, we should further adapt the incremental update scheme into a distributed one. Specifically, a distributed modified IGMM algorithm is first needed for updating the parameters of the GMM-based joint PD. Then, a distributed derivation algorithm is also required to derive the conditional PD from the updated joint PD under the current wind power forecast value of all DWGs. The distributed point-wise incremental learning algorithm and distributed derivation algorithm compose a complete distributed incremental update scheme. 

To the best of our knowledge, neither the distributed point-wise incremental learning algorithm, the distributed derivation algorithm, or the final distributed incremental update scheme have previously been reported. Therefore, in this paper, we aim to propose this distributed incremental update scheme to update each MP's conditional PD of WPFE. The original contributions of this paper are as follows:

\begin{itemize}
\item We propose a distributed framework for the modified IGMM algorithm. Based on this framework, we further develop a  distributed modified IGMM algorithm. This algorithm can enable each MP to efficiently update the parameters of the joint PD when new data arrives.
\item We propose a distributed framework for the derivation process of the conditional PD. According to this framework, we further design a distributed derivation algorithm. This algorithm can enable each MP to derive the conditional PDs of WPFE under all DWGs' latest forecast values from the joint PD.
\item We develop a distributed incremental update scheme by combining the above two original algorithms. This scheme is fully distributed and can keep each MP's conditional PDs of WPFE stay up to date with extremely low and stable update costs regardless of how much the historical dataset grows.
\end{itemize}

The remainder of this paper is organized as follows. In Section 2, the GMM-based joint PD and conditional PD of the WPFE are provided. The distributed framework for the derivation process of the conditional PD is formulated in Section 3. In Section 4, the distributed framework for the modified IGMM algorithm is formulated. The distributed modified IGMM algorithm, the distributed derivation algorithm, and the distributed incremental update scheme are proposed in Section 5. Case studies are performed in Section 6. Finally, Section 7 concludes this paper.

\vspace{10pt}
\noindent \textbf{2. GMM-based Probability Distribution} 
\vspace{5pt}

In this section, the joint PD for correlated DWGs is first introduced. Then, the derivation for the conditional PD of the WPFE from the joint PD is presented. 

\vspace{5pt}
\noindent \textbf{2.1 Joint Probability Distribution} 
\vspace{5pt}

We assume that the number of MPs is \textit{M}. Besides, each MP owns one DWG. Denote the wind power, wind power forecast and WPFE of the MPs by random vectors $\boldsymbol{X}, \boldsymbol{Y}, \boldsymbol{Z} \in \mathbb{R}^{M}$, respectively, where $\boldsymbol{Z} = \boldsymbol{X} - \boldsymbol{Y}$. The \textit{m}-th elements of $\boldsymbol{X}$, $\boldsymbol{Y}$ and $\boldsymbol{Z}$, i.e., $\boldsymbol{x}_m$,  $\boldsymbol{y}_m$ and $\boldsymbol{z}_m$, represent the wind power, wind power forecast and WPFE of the \textit{m}-th MP. The GMM-based joint PD of $\boldsymbol{X}$ and $\boldsymbol{Y}$ with \textit{J} Gaussian components is given in (\ref{Joint GMM}):
\begin{align}
	& P({\mathbf{X}},{\mathbf{Y}})=\sum_{j=1}^J w_j 
	\mathcal N_j({\mathbf{X}},{\mathbf{Y}}) = \sum_{j=1}^J w_j 
	\mathcal N_j(\mathbf{U};\boldsymbol{\mu}_j, \boldsymbol{\Sigma}_j)  \label{Joint GMM} \\
	& \mathcal N_j(\mathbf{U};\boldsymbol{\mu}_j, \boldsymbol{\Sigma}_j)  =  
     \frac{exp[-\frac{1}{2} (\mathbf{U}-\boldsymbol{\mu}_j) \boldsymbol{\Sigma}_j^{-1} (\mathbf{U}-\boldsymbol{\mu}_j)^T]}{\sqrt{(2\pi)^{2M} det(\boldsymbol{\Sigma}_j)}} \label{Gaussian distribution} 
\end{align}
where $\mathcal N_j(\cdot)$ is the \textit{j}-th \textit{2M}-dimensional Gaussian distribution as in (\ref{Gaussian distribution}) with a weighted coefficient $w_j$, a mean vector $\boldsymbol{\mu}_j \in \mathbb{R}^{2M}$ and a covariance matrix $\boldsymbol{\Sigma}_j \in \mathbb{R}^{2M \times 2M}$. $\boldsymbol{U}$ is the compact form of $[\boldsymbol{X},\boldsymbol{Y}]$. The parameter set of the joint PD is defined as $\boldsymbol{\theta}=\{w_j,\boldsymbol{\mu}_j,\boldsymbol{\Sigma}_j|j=1,2,...,J\}$, where the details of $\boldsymbol{\mu}_j$ and $\boldsymbol{\Sigma}_j$ are provided in (\ref{mu}) -- (\ref{C}). 
\begin{align}
    & \boldsymbol{\mu}_j = 
      \begin{bmatrix}
      	  \boldsymbol{\mu}_{j,x} & \boldsymbol{\mu}_{j,y}
      \end{bmatrix}, \ \boldsymbol{\mu}_{j,x} =
	      \begin{bmatrix}
	      	  \mu_{j,x_1} & \cdots & \mu_{j,x_M} 
	      \end{bmatrix} \label{mu} \\
	&  \boldsymbol{\mu}_{j,y} =
	      \begin{bmatrix}
	      	  \mu_{j,y_1} & \cdots & \mu_{j,y_M} 
	      \end{bmatrix}, \ \boldsymbol{\Sigma}_j =
    		\begin{bmatrix}
				\boldsymbol{A}_j  & \boldsymbol{B}_j\\
				\boldsymbol{B}_j^T  & \boldsymbol{C}_j
			\end{bmatrix} \label{mu and sigma} \\
	&  \boldsymbol{A}_j =
    		\begin{bmatrix}
				\sigma_{j,x_1,x_1}  & \cdots & \sigma_{j,x_1,x_M} \\
				\vdots              & \ddots & \vdots             \\
				\sigma_{j,x_M,x_1}  & \cdots & \sigma_{j,x_M,x_M} 
			\end{bmatrix} \label{A} \\
	&  \boldsymbol{B}_j = 
			\begin{bmatrix}
				\sigma_{j,x_1,y_1}  & \cdots & \sigma_{j,x_1,y_M} \\
				\vdots              & \ddots & \vdots             \\
				\sigma_{j,x_M,y_1}  & \cdots & \sigma_{j,x_M,y_M} 
			\end{bmatrix} = 
			\begin{bmatrix}
				\boldsymbol{b}_{j,1} \\
				\vdots \\
				\boldsymbol{b}_{j,M} 
			\end{bmatrix} \label{B} \\
	& \boldsymbol{C}_j =  	
			\begin{bmatrix}
				\sigma_{j,y_1,y_1}  & \cdots & \sigma_{j,y_1,y_M} \\
				\vdots              & \ddots & \vdots             \\
				\sigma_{j,y_M,y_1}  & \cdots & \sigma_{j,y_M,y_M} 
			\end{bmatrix}  \label{C} \\
	& \boldsymbol{C}_j^{-1}  = 
	 		\begin{bmatrix}
				\rho_{j,y_1,y_1} 	& \cdots & \rho_{j,y_1,y_M}  \\
				\vdots              & \ddots & \vdots             \\
				\rho_{j,y_M,y_1}  & \cdots & \rho_{j,y_M,y_M} 
			\end{bmatrix}  \label{C-1}		
\end{align}

Please note that the old joint PD can be achieved by all MPs using the distributed method proposed in \cite{jia2018privacy}. 

\vspace{5pt}
\noindent \textbf{2.2 Derivation of the Conditional Probability Distribution} 
\vspace{5pt}

Once the joint PD in (\ref{Joint GMM}) is obtained, each MP's conditional PD of the WPFE under a given wind power forecast $\boldsymbol{y}^0 \in \mathbb{R}^{M}$ can be derived from the joint PD according to the conditional probability invariance characteristic of the GMM \cite{wang2018}. Define set $\mathcal{M} = \{1,...,M\}$, and the details of the $m$-th MP's ($\forall m \in \mathcal{M}$) conditional PD are shown in (\ref{Conditional GMM}), 
\begin{align}
	& P(z_m|\boldsymbol{y}^0) = \sum\nolimits_{j=1}^J \alpha_{j,m} 
	\mathcal N_j(z_m+y_m|\boldsymbol{y}^0;\lambda_{j,m},\Delta_{j,m})  \label{Conditional GMM} 
\end{align}
where its weighted coefficient $\alpha_{j,m}$ is given in (\ref{alpha}), its mean $\lambda_{j,m}$ is given in (\ref{lambda}) and its variance $\Delta_{j,m}$ is given in (\ref{Delta}). 
\begin{align}
	& \alpha_{j,m} = \frac{w_j \mathcal N_j(\boldsymbol{y}^0;\boldsymbol{\mu}_{j,y}, \boldsymbol{C}_j)}{\sum_{j=1}^J w_j 
	\mathcal N_j(\boldsymbol{y}^0;\boldsymbol{\mu}_{j,y},\boldsymbol{C}_j)} \label{alpha} \\[2mm]
	& \lambda_{j,m} = \mu_{j,x_m} + \boldsymbol{b}_{j,m} \boldsymbol{C}_j^{-1} (\boldsymbol{y}^0-\boldsymbol{\mu}_{j,y}) \label{lambda} \\[2mm]
	& \Delta_{j,m} = \sigma_{j,x_m,x_m} - \boldsymbol{b}_{j,m} \boldsymbol{C}_j^{-1} \boldsymbol{b}_{j,m}^T \label{Delta}
\end{align}
These parameters form a parameter set $\boldsymbol{\gamma}_{j,m} =\{\alpha_{j,m}, \lambda_{j,m}, \Delta_{j,m}\}$. Because the conditional PD in this paper always refers to the conditional PD of the WPFE, we will hereinafter omit `WPFE' for simplicity.
\vspace{10pt}
\noindent \textbf{3. Distributed Framework for the Conditional PD Derivation} 
\vspace{5pt}

Although we have to first update the parameters of the joint PD and then derive the latest conditional PD of each MP using (\ref{alpha})-(\ref{Delta}) under the given $\boldsymbol{y}^0$, a better understanding of the conditional PD derivation can tell us which parameters in the joint PD we must update. Therefore, we investigate the conditional PD derivation before the joint PD update. More specifically, this section formulates a distributed framework for the conditional PD derivation. This framework splits the derivation into local and global parts. The local part can be directly calculated by each MP with its own data, and the global part requires the data of all MPs. The significance of this framework are twofold: 

\begin{itemize}
	\item The global part is actually the calculation part that involves the data of all MPs when deriving the conditional PD. How to deal with the global part by utilizing the information of the local part is the key to developing the distributed derivation algorithm, and thus the clarification of the global part is necessary; 
	\item The local part must be calculated by each MP with its own data. Since some parameters from the updated joint PD are involved in the local part, each MP must update and obtain those parameters to perform its local calculation. Thus, the clarification of the local part can tell us which parameters each MP must update, indicating the direction for the following design of the distributed modified IGMM.
\end{itemize}

\vspace{5pt}
\noindent \textbf{3.1 Distributed Framework for the Derivation of $\alpha_{j,m}$} 
\vspace{5pt}

The derivation of $\alpha_{j,m}$ requires the raw forecast data of all MPs. Thus, we formulate this derivation into local and global parts. First of all, we use superscripts `$new$' to represent the updated parameters. In fact, $\alpha_{j,m}^{new}$ can be written as a function of $d_M^2(\boldsymbol{y}^0,j)$ in (\ref{alpha new}), 
\begin{align}
	& \alpha_{j,m}^{new} = f(d_M^2(\boldsymbol{y}^0,j))  =  \frac{\eta_j w_j^{new} exp[-\frac{1}{2}d_M^2(\boldsymbol{y}^0,j)]}{\sum_{j=1}^J \eta_j w_j^{new} exp[-\frac{1}{2}d_M^2(\boldsymbol{y}^0,j)]} \label{alpha new} \\[2mm]
	& \eta_j = \sqrt{(2\pi)^{2M}det(\boldsymbol{C}_j^{new})} \notag 
\end{align}
where $d_M^2(\boldsymbol{y}^0,j)$ is the squared Mahalanobis distance between the \textit{j}-th Gaussian component and $\boldsymbol{y}^0$, as given in (\ref{d_M y0}). 
\begin{align}
	& d_M^2(\boldsymbol{y}^0,j) = (\boldsymbol{y}^0-\boldsymbol{\mu}_{j,y}^{new})\boldsymbol{C}_j^{new^{-1}} (\boldsymbol{y}^0-\boldsymbol{\mu}_{j,y}^{new})^T \label{d_M y0}
\end{align}

Except for obtaining $w_j^{new}$ and $\boldsymbol{C}_j^{new}$, each MP still needs $\boldsymbol{y}^0$ and $\boldsymbol{\mu}_{j,y}^{new}$ to calculate $d_M^2(\boldsymbol{y}^0,j)$. To avoid collecting $\boldsymbol{y}^0$, we formulate the calculation of (\ref{d_M y0}) into local and global parts, as given in (\ref{local alpha 1})-(\ref{global alpha 2}). Equation (\ref{local alpha 1}) and (\ref{local alpha 2}) are the local parts, while (\ref{global alpha 1}) and (\ref{global alpha 2}) are the global parts. 
\begin{align}
	& \varrho_{j,m,i}  =  \rho_{j,y_m,y_i}^{new} \left(y_m^0 - \mu_{j,y_m}^{new} \right) \label{local alpha 1} \\[2mm]
	& \vartheta_{j,i} = \sum_{m=1}^M \varrho_{j,m,i} \label{global alpha 1} \\[2mm]
	& \varphi_{j,i}(\vartheta_{j,i}) = \vartheta_{j,i} \left(y_i^0 - \mu_{j,y_i}^{new} \right) \label{local alpha 2}\\[2mm]
	& d_M^2(\boldsymbol{y}^0,j) = \sum_{i=1}^M \varphi_{j,i}(\vartheta_{j,i}) \label{global alpha 2}
\end{align}

Equation (\ref{local alpha 1})-(\ref{global alpha 2}) should be calculated sequentially. Specifically, the $m$-th MP ($\forall m \in \mathcal{M}$) needs to compute $\varrho_{j,m,i}$ in (\ref{local alpha 1}) with its own data as well as $\mu_{j,y_m}^{new}$ and $\rho_{j,y_m,y_i}^{new}$, where $\rho_{j,y_m,y_i}^{new}$ is the updated elements in (\ref{C-1}). To calculate $\vartheta_{j,i}$ in (\ref{global alpha 1}), each MP is required to collect the calculation results $\varrho_{j,m,i}$ of the others. After obtaining $\vartheta_{j,i}$, the $i$-th MP ($\forall i \in \mathcal{M}$) needs to compute $\varphi_{j,i}$ in (\ref{local alpha 2}) by its own data and $\mu_{j,y_i}^{new}$. To finally compute $d_M^2(\boldsymbol{y}^0,j)$ in (\ref{global alpha 2}), each MP still needs to collect the calculation results $\varphi_{j,i}$ of the others.  

Note that the relationship between (\ref{local alpha 1}) and (\ref{global alpha 1}) and the relationship between (\ref{local alpha 2}) and (\ref{global alpha 2}) are essentially the same, i.e., the global part is the sum of the local parts. This relationship can be abstracted into the following, where $L_m$ is the local calculation result of the $m$-th MP ($\forall m \in \mathcal{M}$) and $G$ is the global information that should be obtained by each MP. 
\begin{equation}
	G = \sum_{m=1}^M L_m \label{relationship J}
\end{equation}

Therefore, the key to derive $\alpha_{j,m}^{new}$ in (\ref{alpha new}) lies in finding a distributed method to obtain the global summation among the local calculation results of the MPs. Meanwhile, the $m$-th MP ($\forall m \in \mathcal{M}$) should update and obtain $w_j^{new}$, $\boldsymbol{C}_j^{new}$ and $\mu_{j,y_m}^{new}$ to perform its local calculation.

\vspace{5pt}
\noindent \textbf{3.2 Distributed Framework for the Derivation of $\lambda_{j,m}$} 
\vspace{5pt}

The derivation of $\lambda_{j,m}^{new}$ requires the raw forecast data of all MPs, and thus we will formulate it into the local and global parts. First of all, we rewrite the derivation of $\lambda_{j,m}^{new}$ into (\ref{new lambda}), where $\vartheta_{j,i}$ is given in (\ref{global alpha 1}) and $\sigma_{j,x_m,y_i}$ is the element in (\ref{B}).
\begin{align}
	& \lambda_{j,m}^{new} = \mu_{j,x_m}^{new} + \sum_{i=1}^M \sigma_{j,x_m,y_i}^{new} \vartheta_{j,i} \label{new lambda}
\end{align}

To achieve $\lambda_{j,m}^{new}$, the $m$-th MP ($\forall m \in \mathcal{M}$) should first compute (\ref{local alpha 1}) using its own data as well as $\rho_{j,y_m,y_i}^{new}$ and $\mu_{j,y_m}^{new}$. Then, the MPs need to cooperate to calculate (\ref{global alpha 1}) to achieve $\vartheta_{j,i}$. Thereafter, the $m$-th MP ($\forall m \in \mathcal{M}$) can compute $\lambda_{j,m}^{new}$ in (\ref{new lambda}) using $\mu_{j,x_m}^{new}$ in (\ref{mu}) and $\boldsymbol{b}_{j,m}^{new}$ in (\ref{B}). Thus, under the premise of obtaining $\rho_{j,y_m,y_i}^{new}$, $\mu_{j,y_m}^{new}$, $\mu_{j,x_m}^{new}$ and $\boldsymbol{b}_{j,m}^{new}$ by the $m$-th MP, the local part of deriving $\lambda_{j,m}^{new}$ is the calculation of (\ref{local alpha 1}), while its global part is the calculation of (\ref{global alpha 1}). Note that the relationship between the local and global parts can also be abstracted into (\ref{relationship J}). 

Therefore, the key to deriving $\lambda_{j,m}^{new}$ also lies in finding a distributed method to achieve the global summation of the local calculation results of the MPs. Meanwhile, the $m$-th MP ($\forall m \in \mathcal{M}$) should update and obtain $\boldsymbol{C}_j^{new}$, $\mu_{j,y_m}^{new}$, $\mu_{j,x_m}^{new}$ and $\boldsymbol{b}_{j,m}^{new}$ to perform its local calculation.

\vspace{5pt}
\noindent \textbf{3.3 Distributed Framework for the Derivation of $\Delta_{j,m}$} 
\vspace{5pt}

The derivation of $\Delta_{j,m}^{new}$ does not require the data of other MPs. Thus, once $\sigma_{j,x_m,x_m}^{new}$, $\boldsymbol{C}_j^{new}$ and $\boldsymbol{b}_{j,m}^{new}$ are obtained by the $m$-th MP ($\forall m \in \mathcal{M}$), it can perform the derivation of $\Delta_{j,m}^{new}$ via a completely local calculation.

\vspace{10pt}
\textit{\textbf{Remark 1}:} The key to deriving $\alpha_{j,m}^{new}$ and  $\lambda_{j,m}^{new}$ lies in finding a distributed manner to obtain the global summation of the local calculation results of MPs, as given in (\ref{relationship J}). Deriving $\Delta_{j,m}^{new}$ is a completely local calculation without the need for cooperation with the other MPs. 
\vspace{10pt}

\textit{\textbf{Remark 2}:} The $m$-th MP ($\forall m \in \mathcal{M}$) should update and obtain $w_j^{new}$, $\mu_{j,x_m}^{new}$, $\mu_{j,y_m}^{new}$, $\sigma_{j,x_m,x_m}^{new}$, $\boldsymbol{C}_j^{new}$ and $\boldsymbol{b}_{j,m}^{new}$ to perform its local calculations of the whole derivation for its conditional PD.

\vspace{10pt}
\noindent \textbf{4. Distributed Framework for the Modified Incremental GMM} 
\vspace{5pt}

From the previous section, we already know which parameters in the joint PD each MP should update. In this section, we first briefly review the modified IGMM algorithm \cite{pinto2015fast} to show how to update those parameters in an incremental but centralized manner. Then, we formulate a distributed framework for the modified IGMM by splitting the algorithm into local and global parts. This distributed framework can help us to look for the key to designing the distributed modified IGMM.

\vspace{5pt}
\noindent \textbf{4.1 Modified IGMM Algorithm} 
\vspace{5pt}

The modified IGMM consists of three steps: the judging step, the updating step, and the creating step. 

The judging step is to evaluate whether a piece of new data obeys the existing old GMM. Given a piece of new data $\boldsymbol{u} = [\boldsymbol{u}_x, \boldsymbol{u}_y]\in \mathbb{R}^{2M}$, the \textit{m}-th and the (\textit{M+m})-th elements of $\boldsymbol{u}$ are the new wind power and wind power forecast data of the \textit{m}-th MP, represented as $u_{x_m}$ and $u_{y_m}$, respectively. The judging step is presented in (\ref{judging}) based on the squared Mahalanobis distance $d_M^2(\boldsymbol{u},j)$ between the \textit{j}-th component and $\boldsymbol{u}$, where the superscripts `$old$' represent the parameters before the update. $\chi_{D,1-\beta}^2$ is the $1-\beta$ percentile of a chi-squared distribution with \textit{D} degrees of freedom.
\begin{align}
	d_M^2(\boldsymbol{u},j) & = (\boldsymbol{u}-\boldsymbol{\mu}_j^{old})\boldsymbol{\Sigma}_j^{old^{-1}} (\boldsymbol{u}-\boldsymbol{\mu}_j^{old})^T \leq \chi_{D,1-\beta}^2, \, \, j=1,...,J \label{judging} 
\end{align}

For the updating step, if (\ref{judging}) holds, $\boldsymbol{\theta}_{j,m}^{new}$ will be updated based on the Robbins-Monro stochastic approximation, as given in (\ref{new w})-(\ref{new sigma}), where $w_j$ is also the prior probability of the $j$-th component of the GMM.
\begin{align}
	& w_j^{new} = h_j^{new}/\sum\nolimits_{j=1}^J h_j^{new} \label{new w} \\[1mm]
	& \mu_{j,x_m}^{new} = \mu_{j,x_m}^{old} + r_j(u_{x_m}-\mu_{j,x_m}^{old}) \label{new mu x} \\[2mm]
	& \mu_{j,y_m}^{new} = \mu_{j,y_m}^{old} + r_j(u_{y_m}-\mu_{j,y_m}^{old}) \label{new mu y} \\[2mm]
	& \boldsymbol{\Sigma}_j^{new} = \left(1-r_j \right) \boldsymbol{\Sigma}_j^{old}+r_j\boldsymbol{e}_j^T\boldsymbol{e}_j-\Delta\boldsymbol{\mu}_j^T\Delta\boldsymbol{\mu}_j \label{new sigma} 
\end{align} 
Auxiliary parameters are given as follows, where $h_j$ is the accumulator of the posterior probability and $p(j|\boldsymbol{u})$ is the posterior probability for the \textit{j}-th component.
\begin{align}
	& h_j^{new} = h_j^{old} + p(j|\boldsymbol{u}),\ \ r_j = p(j|\boldsymbol{u}) / h_j^{new} \notag \\[2mm]
	& \boldsymbol{e}_j = \boldsymbol{u}-\boldsymbol{\mu}_j^{new},\ \ \Delta\boldsymbol{\mu}_j = r_j \left(\boldsymbol{u}-\boldsymbol{\mu}_j^{old}\right) \notag \\[1mm]
	& p(j|\boldsymbol{u}) = \frac{w_j^{old} \mathcal N_j(\boldsymbol{u};\boldsymbol{\mu}_{j}^{old}, \boldsymbol{\Sigma}_j^{old})}{\sum_{j=1}^J w_j^{old} 
	\mathcal N_j(\boldsymbol{u};\boldsymbol{\mu}_{j}^{old}, \boldsymbol{\Sigma}_j^{old})}  \notag
\end{align}

For the creating step, if (\ref{judging}) does not hold, then a new Gaussian component should be created to accommodate the new information carried by the new data. The $\boldsymbol{\theta}_{j,m}^{new}$ of this new component are initialized by (\ref{create w})-(\ref{create sigma}), where $\boldsymbol{\Delta}_{ini}$ is a preset parameter. 
\begin{align}
	& w_{J+1}= \left. 1 \middle/ \right( 1+\sum\nolimits_{j=1}^{J} h_j^{old} \left) \right.\label{create w} \\
	& \mu_{J+1,x_m} = u_{x_m} \label{create mu x}\\[1mm]
	& \mu_{J+1,y_m} = u_{y_m} \label{create mu y}\\[1mm]
	& \boldsymbol{\Sigma}_{J+1} = \boldsymbol{\Delta}_{ini} \label{create sigma} 
\end{align}

\vspace{5pt}
\noindent \textbf{4.2 Distributed Framework for the Judging Step} 
\vspace{5pt}

Every MP needs to obtain $d_M^2(\boldsymbol{u},j)$ to perform the judgment in (\ref{judging}). To formulate the calculation of (\ref{judging}) into local and global parts, we first divide the calculation of $d_M^2(\boldsymbol{u},j)$ into two layers: the first layer is given in (\ref{1st layer global x}) and (\ref{1st layer global y}), and the second layer is given in (\ref{2nd layer global}). These two layers should be calculated sequentially. 
\begin{align}
	& g_{j,x_i} = \sum_{m=1}^M l_{j,m,x_i},\ \ \forall i \in \mathcal{M} \label{1st layer global x} \\
	& g_{j,y_i} = \sum_{m=1}^M l_{j,m,y_i},\ \ \forall i \in \mathcal{M} \label{1st layer global y} \\
	& d_M^2(\boldsymbol{u},j) = \sum_{i=1}^M L_{j,i}(g_{j,x_i},g_{j,y_i}) \label{2nd layer global}
\end{align}

In the first layer, the $m$-th MP ($\forall m \in \mathcal{M}$) should compute $l_{m,x_i}$ in (\ref{1st layer local x}) and $l_{m,y_i}$ in (\ref{1st layer local y}) by its own data as well as the elements of the inverse of $\boldsymbol{\Sigma}_j^{old}$ in (\ref{inverse covariation}). Thus, instead of the elements $\sigma_{j,x_m,x_m}$, $\boldsymbol{C}_j$, and $\boldsymbol{b}_{j,m}$ mentioned in Section 3, the $m$-th MP actually needs to update and obtain the whole $\boldsymbol{\Sigma}_j$. 

\begin{align}
	& l_{j,m,x_i} = \rho_{j,x_m,x_i}^{old}(u_{x_m}-\mu_{x_m}^{old}) + \rho_{j,y_m,x_i}^{old}(u_{y_m}-\mu_{y_m}^{old}),\  \forall i \in \mathcal{M} \label{1st layer local x} \\[2mm] 
	& l_{j,m,y_i} = \rho_{j,x_m,y_i}^{old}(u_{x_m}-\mu_{x_m}^{old}) + \rho_{j,y_m,y_i}^{old}(u_{y_m}-\mu_{y_m}^{old}),\  \forall i \in \mathcal{M} \label{1st layer local y} \\[2mm] 
	& \boldsymbol{\Sigma}_j^{old^{-1}}  = 
	 		\begin{bmatrix}
				\rho_{j,x_1,x_1}^{old}  	& \cdots & \rho_{j,x_1,y_M}^{old}  \\
				\vdots              & \vdots & \vdots             \\
				\rho_{j,y_M,x_1}^{old}  & \cdots & \rho_{j,y_M,y_M}^{old} 
			\end{bmatrix} \label{inverse covariation}
\end{align} 

\vspace{10pt}
\textit{\textbf{Remark 3}:} Combining the analysis under the reformulation of deriving the conditional PD and the judging step, we finally come to the conclusion that the $m$-th MP ($\forall m \in \mathcal{M}$) should update and obtain the parameters $\boldsymbol{\theta}_{j,m}^{new} = \{w_j^{new}, \mu_{j,x_m}^{new}, \mu_{j,y_m}^{new}, \boldsymbol{\Sigma}_j^{new}|\  j=1,...,J\}$ for the final update of the conditional PD. Besides, the $m$-th MP knows $\boldsymbol{\theta}_{j,m}^{old}$ after the last update.
\vspace{10pt}

However, to calculate $g_{x_i}$ in (\ref{1st layer global x}) and $g_{y_i}$ in (\ref{1st layer global y}), each MP is required to collect the calculation results of the other MPs. Thus, the calculations of (\ref{1st layer local x}) and (\ref{1st layer local y}) are the local parts in the first layer of the judging step, and the calculations of (\ref{1st layer global x}) and (\ref{1st layer global y}) are the corresponding global parts, which require the cooperation of the MPs.

If each MP has completed the calculation in the first layer and obtained $g_{x_i}$ and $g_{y_i}$, the $i$-th MP ($\forall i \in \mathcal{M}$) can perform the calculation in (\ref{2nd layer local}) using its own data. However, to calculate $d_M^2(\boldsymbol{u},j)$ in (\ref{2nd layer global}), each MP still requires the calculation results of the other MPs. Thus, the calculation of (\ref{2nd layer local}) is the local part in the second layer of the judging step, and the calculation of (\ref{2nd layer global}) is the corresponding global part that needs the cooperation of the MPs.
\begin{equation}
	L_{j,i}(g_{j,x_i},g_{j,y_i}) = g_{j,x_i}(u_{x_i}-\mu_{x_m}^{old}) + g_{j,y_i}(u_{y_i}-\mu_{y_m}^{old}) \label{2nd layer local}
\end{equation}

Note that the relationship between the local part and the global part in the judging step can also be abstracted into (\ref{relationship J}), whether it is in the first layer or the second layer. Thus, the key to realizing a distributed judging step also lies in finding a distributed method to obtain the global summation of the local calculation results of the MPs. 

\vspace{5pt}
\noindent \textbf{4.2 Distributed Framework for the Updating Step} 
\vspace{5pt}
If (\ref{judging}) holds according to MPs' judgements, the $m$-th MP ($\forall m \in \mathcal{M}$) needs to update $\boldsymbol{\theta}_{j,m}^{new}$. Note that each MP already have the value of $d_M^2(\boldsymbol{u},j)$ after the judging step. Under this premise, we formulate the updating step into local and global parts. 

\vspace{3pt}
\textit{1) Updating $w_j$}
\vspace{3pt}

Each MP only needs to compute $p(j|\boldsymbol{u})$ to obtain $w_j^{new}$ in (\ref{new w}). In fact, $p(j|\boldsymbol{u})$ is a function of $d_M^2(\boldsymbol{u},j)$ as given in (\ref{function d}). Once $d_M^2(\boldsymbol{u},j)$ is obtained by each MP, with the known old parameters $w_j^{old}$ and $\boldsymbol{\Sigma}_j^{old}$, each MP can calculate (\ref{function d}) by itself. Therefore, the update of $w_j$ is a completely local calculation without requiring the cooperation of MPs.
\begin{align}
	p(j|\boldsymbol{u}) & = \frac{c_j w_j^{old} exp[-\frac{1}{2}d_M^2(\boldsymbol{u},j)]}{\sum_{j=1}^J c_j w_j^{old} exp[-\frac{1}{2}d_M^2(\boldsymbol{u},j)]} \label{function d} \\[2mm]
	c_j & = \sqrt{(2\pi)^{2M}det(\boldsymbol{\Sigma}_j^{old})} \notag
\end{align}

\vspace{3pt}
\textit{2) Updating $\mu_{j,x_m}$ and $\mu_{j,y_m}$}
\vspace{3pt}

The $m$-th MP ($\forall m \in \mathcal{M}$) is able to update $\mu_{j,x_m}$ in (\ref{new mu x}) and $\mu_{j,y_m}$ in (\ref{new mu y}) using its own data, the old parameters, and the known $r_j$ after obtaining $p(j|\boldsymbol{u})$ by $d_M^2(\boldsymbol{u},j)$. Therefore, the updates of $\mu_{j,x_m}$ and $\mu_{j,y_m}$ are still completely local calculations without requiring the cooperation of MPs.

\vspace{3pt}
\textit{3) Updating $\boldsymbol{\Sigma}_j$}
\vspace{3pt}

We first define the element of $\boldsymbol{\Sigma}_j$ in the \textit{m}-th row and the \textit{i}-th column as $\sigma_{j,v_m,v_i}$, where the subscript $v_m$ represents the subscript $x_m$ if $m \leq M$; otherwise, it represents the subscript $y_{m-M}$. Subscript $v_i$ is the same. The update of $\sigma_{j,v_m,v_i}$ is rewritten into (\ref{updated sigma}). For the $m$-th MP ($\forall m \in \mathcal{M}$), its local calculations are given in (\ref{local sigma 1})-(\ref{local sigma 3}) after obtaining $p(j|\boldsymbol{u})$ by $d_M^2(\boldsymbol{u},j)$. However, to update $\sigma_{j,v_m,v_i}$ in (\ref{updated sigma}), the $m$-th MP ($\forall m \in \mathcal{M}$) still needs to collect the local calculation result $\xi_{i}$ from the $i$-th MP ($\forall i \in \mathcal{M}$), which is given in (\ref{local sigma 4}). Once a completed vector $\boldsymbol{\xi}$ in (\ref{global sigma}) is collected by the $m$-th MP, it can finally achieve the update of $\boldsymbol{\Sigma}_j$. Therefore, the local parts of updating $\boldsymbol{\Sigma}_j$ are the calculations of (\ref{local sigma 1})-(\ref{local sigma 3}), and the corresponding global part is to form $\boldsymbol{\xi}$ in (\ref{global sigma}) by collecting the local calculation results of the other MPs.
\begin{align}
	& \sigma_{j,v_m,v_i}^{new}  = \epsilon_{m,i}+\varepsilon \xi_{m}\xi_{i}, \ \ \forall m,i \in \{1,...,2M\} \label{updated sigma} \\[1mm]
	& \varepsilon = r_j(1+r_j^2-3r_j) \label{local sigma 1} \\[1mm]
	& \epsilon_{m,i} = (1-r_j)\sigma_{j,v_m,v_i}^{old}, \ \ \forall m,i \in \{1,...,2M\} \label{local sigma 2} \\[1mm]
	&  \xi_{m} = u_{v_m}-\mu_{j,v_m}^{old} \ \label{local sigma 3} \\[1mm]
	& \xi_{i}  = u_{v_i}-\mu_{j,v_i}^{old} \label{local sigma 4} \\[1mm]
	& \boldsymbol{\xi}  = [\xi_1\ \cdots \ \xi_{M},\ \xi_{M+1}\ \cdots \ \xi_{M+M}] \label{global sigma}
\end{align}

\vspace{3pt}
\textit{4) The relationship between the local and global parts in the updating step}
\vspace{3pt}

In fact, the global part of the updating step, i.e., collecting the local calculation results of the other MPs to form (\ref{global sigma}), can be abstracted into the following:
\begin{equation}
	\boldsymbol{Q} = [R_1\ \cdots \ R_m\ \cdots \ R_{m+M}\ \cdots \  R_{2M}] \label{relationship U}
\end{equation}
\noindent where $R_m$ and $R_{m+M}$ are the local calculation results of the $m$-th MP ($\forall m \in \mathcal{M}$). $\boldsymbol{Q}$ should be obtained by all MPs. Therefore, we know that the key to developing a distributed updating step lies in finding a distributed method to collect the local calculation results of the MPs.

\vspace{5pt}
\noindent \textbf{4.3 Distributed Framework for the Creating Step} 
\vspace{5pt}

If (\ref{judging}) does not hold according to MPs' judgements, every MP needs to create a new component. First, with the known old accumulator $h_j^{old}$, the $m$-th MP ($\forall m \in \mathcal{M}$) can directly create $w_{J+1}$ by \eqref{create w}. Second, with its own data, the $m$-th MP can also create $\mu_{J+1,x_m}$ and $\mu_{J+1,y_m}$ by \eqref{create mu x} and \eqref{create mu y}, respectively. Finally, as $\boldsymbol{\Delta}_{ini}$ is a preset parameter known by every MP, the $m$-th MP can directly create $\boldsymbol{\Sigma}_{J+1}$. Therefore, the creating step is a completely local calculation for each MP without the need of the data from other MPs.

\vspace{10pt}
\textit{\textbf{Remark 4}:} The key to performing the distributed judging step lies in finding a distributed method to obtain the global summation of the local calculation results of the MPs, as given in (\ref{relationship J}), while the key to performing the distributed updating step lies in finding a distributed method to collect the local calculation results of the MPs, as given in (\ref{relationship U}). Performing the creating step only requires each MP's local calculation.
\vspace{10pt}

\vspace{10pt}
\noindent \textbf{5. Distributed Incremental Update Scheme} 
\vspace{5pt}

In this section, we first use the average consensus algorithm to deal with the keys revealed in Section 3 and 4. After that, we propose a distributed modified IGMM algorithm. Then, we develop a distributed derivation algorithm. Hereafter, a completed distributed incremental update scheme is obtained by combining the two original distributed algorithms. Finally, the features of the proposed scheme are analyzed and discussed. 


\vspace{5pt}
\noindent \textbf{5.1 Realization of the Global Summation and Collection} 
\vspace{5pt}

According to Remark 1 and 4, the keys to designing a distributed derivation algorithm and a distributed modified IGMM lie in: 1) finding a distributed method to obtain the global summation of the local calculation results of the MPs, as given in (\ref{relationship J}); and 2) finding a distributed method to collect the local calculation results of the MPs, as given in (\ref{relationship U}). To deal with the two keys, we use the average consensus algorithm to achieve the global summation and collection.

The average consensus algorithm is a graph-theory-based method. Therefore, we consider \textit{M} MPs as \textit{M} nodes in a connected communication network, where each MP can only communicate with its neighbors. The neighbor set of the \textit{m}-th MP is defined as $\Phi_m$. Denote the degrees of node $m$ as $D_m$. Based on the information exchange between the MPs in each iteration of the average consensus algorithm, every MP can finally achieve the average value of their initial input after convergence. For the derivation and discussion, refer to \cite{4118472}.

First, we use the average consensus algorithm to achieve the global summation in (\ref{relationship J}). The $t$-th iteration of the average consensus algorithm for the $m$-th MP ($\forall m \in \mathcal{M}$) is given in (\ref{ACA 1}):
\begin{align}
	G_m^{t+1} & = G_m^{t} + \sum\nolimits_{i \in \Phi_m}\zeta_{m,i}\left[ G_i^{t}-G_m^{t} \right],\ \ G_m^{0} = L_m \label{ACA 1}
\end{align}
where $G_i^{t}$ is the information received by the $m$-th MP from its neighbors, and $\zeta_{m,i}$ is the adjacency coefficient given as follows: 
\begin{align}
	\zeta_{m,i} & = 2/[D_m+D_i+1],  \ \forall i \in \Phi_m \notag
\end{align}
Once the convergence of the average consensus algorithm is achieved, each MP can obtain the average value in (\ref{average value 1}). Because the value of $M$ is public knowledge, each MP can eventually obtain the result of (\ref{relationship J}) from (\ref{average value 1}).
\begin{align}
	\lim_{t\rightarrow \infty} G_m^t & = \frac{1}{M}\sum\nolimits_{m=1}^ML_m 
	\label{average value 1}
\end{align} 

Then, to achieve the global collection in (\ref{relationship U}) in a distributed manner, we also use the average consensus algorithm but with a certain design of each MP's initial input. The initial input of the $m$-th MP ($\forall m \in \mathcal{M}$) is given in (\ref{initial input 2}), which is a sparse vector only with value in the $m$-th and the $(m+M)$-th elements, while the other elements are 0. Based on this initial value design, each MP can perform the average consensus algorithm in (\ref{ACA 2}) by receiving $\boldsymbol{Q}_i^{t}$ from its neighbors. The converged result of (\ref{ACA 2}) for the $m$-th MP is denoted by $\boldsymbol{Q}_m$ and provided in (\ref{average value 2}). Clearly, the $m$-th MP can obtain $\boldsymbol{Q}$ from (\ref{average value 2}) after the convergence of (\ref{ACA 2}).
\begin{align}
	& \boldsymbol{Q}_m^0 = \left[\ \underbrace{\cdots \ \ \cdots} \ \ R_m \ \ \underbrace{\cdots \ \ \cdots} \ \ R_{m+M}\ \  \underbrace{\cdots \ \ \cdots} \ \ \right] \label{initial input 2} \\
	& \qquad \qquad \quad \,  0 \qquad \qquad \qquad  0 \qquad \qquad \qquad \quad \  0 \notag \\
	& \boldsymbol{Q}_m^{t+1} = \boldsymbol{Q}_m^{t} + \sum\nolimits_{i \in \Phi_m}\zeta_{m,i}\left[ \boldsymbol{Q}_i^{t}-\boldsymbol{Q}_m^{t} \right] \label{ACA 2} \\[3mm]
	& \boldsymbol{Q}_m = \frac{1}{M} [R_1\ \cdots \ R_m\ \cdots \ R_{m+M}\ \cdots \  R_{2M}] =\frac{1}{M} \boldsymbol{Q} \label{average value 2}
\end{align}

\vspace{1pt}
\noindent \textbf{5.2 Distributed Modified IGMM} 
\vspace{1pt}

The idea of the distributed modified IGMM is that, the $m$-th MP ($\forall m \in \mathcal{M}$) first performs its local calculations of the modified IGMM via its own data and the known old parameters. Then, the $m$-th MP ($\forall m \in \mathcal{M}$) uses the average consensus algorithm discussed above to calculate the global parts of the modified IGMM to finally obtain $\boldsymbol{\theta}_{j,m}^{new}$. The detailed distributed modified IGMM for the $m$-th MP ($\forall m \in \mathcal{M}$) to update $\boldsymbol{\theta}_{j,m}^{new}$ is given in Algorithm 1.
\begin{algorithm}
	\label{distributed-Modified}
	\caption{ The distributed modified IGMM}
	\KwIn{new data $\boldsymbol{u}$}
	\vspace{2pt}
    \KwIn{$\{\boldsymbol{\theta}_{j,m}^{old}|j=1,...,J\}$ of the $m$-th MP ($\forall m \in \mathcal{M}$)}
    \vspace{2pt}
    \KwOut{$\{\boldsymbol{\theta}_{j,m}^{new}|j=1,...,J\}$ of the $m$-th MP ($\forall m \in \mathcal{M}$)}  
    \vspace{2pt}  
    \For{$j=1$ to $J$}
	{	
		The $m$-th MP ($\forall m \in \mathcal{M}$) computes (\ref{1st layer local x}) and (\ref{1st layer local y})\;
		Set $L_m = l_{j,m,x_i}$ and $L_m = l_{j,m,y_i}$  sequentially\;
		Set $t=0$\;
		\While{convergence criterion is not met}
    		{
    			The $m$-th MP ($\forall m \in \mathcal{M}$) performs (\ref{ACA 1})\;
    			$t=t+1$\;
			}
			Each MP obtains $g_{j,x_i}$ and $g_{j,y_i}$ by (\ref{average value 1}) sequentially\;
		The $m$-th MP ($\forall m \in \mathcal{M}$) computes (\ref{2nd layer local})\;
		Set $L_m = L_{j,i}(g_{j,x_i},g_{j,y_i})$ and $t=0$\;
		\While{convergence criterion is not met}
    		{
    			The $m$-th MP ($\forall m \in \mathcal{M}$) performs (\ref{ACA 1})\;
    			$t=t+1$\;
			}
		Each MP obtains $d_M^2(\boldsymbol{u},j)$ from (\ref{average value 1})\;
		Each MP performs a judgment in (\ref{judging})\;
	}
	\eIf{(\ref{judging}) holds }
	{
		\For{$j=1$ to $J$}
		{   
			Each MP computes (\ref{local sigma 1})\;
			The $m$-th MP ($\forall m \in \mathcal{M}$) computes (\ref{local sigma 2}), (\ref{local sigma 3})\;
			Set $R_m=\xi_{M}$ and $R_{m+M}=\xi_{m+M}$\;
			The $m$-th MP ($\forall m \in \mathcal{M}$) forms (\ref{initial input 2})\;
			Set $t=0$\;
			\While{convergence criterion is not met}
    		{
    			Each MP calculates (\ref{ACA 2})\;
    			$t=t+1$\;
			}
			Each MP obtains $\boldsymbol{\xi}$ from (\ref{average value 2})\;
			Each MP updates $\boldsymbol{\Sigma}_j^{new}$ by (\ref{updated sigma})\;			
			Each MP computes (\ref{function d}) by $d_M^2(\boldsymbol{u},j)$\;
			Each MP computes $h_j^{new}$ and $r_j$\;
			Each MP updates $w_j^{new}$ by (\ref{new w})\;
			The $m$-th MP ($\forall m \in \mathcal{M}$) updates $\mu_{j,x_m}^{new}$ by (\ref{new mu y})\;
			The $m$-th MP ($\forall m \in \mathcal{M}$) updates $\mu_{j,y_m}^{new}$ by (\ref{new mu y})\;
		}

	}
	{
		$J=J+1$\;
		Each MP creates $w_{J}$ by (\ref{create w})\;
		Each MP creates $\boldsymbol{\Sigma}_J^{new}$ by (\ref{create sigma})\;
		The $m$-th MP ($\forall m \in \mathcal{M}$) creates $\mu_{J,x_m}^{new}$ by (\ref{create mu x})\;
		The $m$-th MP ($\forall m \in \mathcal{M}$) creates $\mu_{J,y_m}^{new}$ by (\ref{create mu y})\;
	}
\end{algorithm}

\vspace{5pt}
\noindent \textbf{5.3 Distributed Derivation Algorithm} 
\vspace{5pt}

The idea of the distributed derivation algorithm is that, based on $\boldsymbol{\theta}_{j,m}^{new}$, the $m$-th MP ($\forall m \in \mathcal{M}$) first completes its local calculations of the conditional PD derivation. Then, the $m$-th MP ($\forall m \in \mathcal{M}$) calculates the global parts of the derivation using the average consensus algorithm to obtain $\boldsymbol{\gamma}_{j,m}$. The detailed distributed derivation algorithm is given in Algorithm 2.

\begin{algorithm}
	\label{distributed derivation algorithm}
	\caption{The distributed derivation algorithm}\nonl
    \KwIn{$y_m^0$ of the $m$-th MP ($\forall m \in \mathcal{M}$)}
    \vspace{2pt}
    \KwIn{$\{\boldsymbol{\theta}_{j,m}^{new}|j=1,...,J\}$ of the $m$-th MP ($\forall m \in \mathcal{M}$)}
    \vspace{2pt}
    \KwOut{$\{\boldsymbol{\gamma}_{j,m}^{new}|j=1,...,J\}$ of the $m$-th MP ($\forall m \in \mathcal{M}$)}
    \For{$j=1$ to $J$}
	{
		The $m$-th MP ($\forall m \in \mathcal{M}$) computes (\ref{local alpha 1})\;
		Set $L_m = \varrho_{j,m,i}$ and $t=0$\;
		\While{convergence criterion is not met}
		{
			Each MP calculates (\ref{ACA 1})\;
			$t=t+1$\;
		}
		Each MP obtains $\vartheta_{j,i}$ by (\ref{average value 1})\;
		The $m$-th MP ($\forall m \in \mathcal{M}$) computes (\ref{local alpha 2})\;
		Set $L_m = \varphi_{j,i}$ and $t=0$\;
		\While{convergence criterion is not met}
		{
			Each MP calculates (\ref{ACA 1})\;
			$t=t+1$\;
		}
		Each MP obtains $d_M^2(\boldsymbol{y}^0,j)$ by (\ref{average value 1})\;
		The $m$-th MP ($\forall m \in \mathcal{M}$) obtains $\alpha_{j,m}^{new}$ by (\ref{alpha new})\;
		The $m$-th MP ($\forall m \in \mathcal{M}$) obtains $\lambda_{j,m}^{new}$ by (\ref{new lambda})\;
		The $m$-th MP ($\forall m \in \mathcal{M}$) obtains $\Delta_{j,m}^{new}$ by (\ref{Delta})\;
	}
\end{algorithm}

\vspace{5pt}
\noindent \textbf{5.4 Distributed Incremental Update Scheme and Feature Analysis} 
\vspace{5pt}

The complete idea behind the distributed incremental update scheme lies in that, the $m$-th MP ($\forall m \in \mathcal{M}$) should first update the desired parameters using the distributed modified IGMM, and then derive its conditional PD via the distributed derivation algorithm. Therefore, combining the distributed modified IGMM and derivation algorithm, we finally achieve the distributed incremental update scheme. There are four features of the proposed distributed incremental update scheme.

\vspace{3pt}
\textit{1) Stay up to date}
\vspace{3pt}

The distributed incremental update scheme enables each MP to update its conditional PD consistent with all sequentially presented data and keep it up to date with high accuracy.

\vspace{3pt}
\textit{2) Highly efficient}
\vspace{3pt}

The distributed incremental update scheme enables each MP to update its conditional PD in an incremental manner, only needing to perform an update calculation of a piece of new data. Compared with the traditional EM algorithm that needs to consistently reconstruct and retrain the entire historical dataset, this scheme is considerably more efficient with an extremely low and stable update cost regardless of how much the historical dataset grows.

\vspace{3pt}
\textit{3) Fully distributed}
\vspace{3pt}

In the distributed incremental update scheme, each MP only needs to communicate with its neighbors to realize the entire update process. No data center or coordinator is needed. Meanwhile, the correlation between those correlated DWGs is involved in the updated results, leading to a very concentrated conditional PD. This conditional PD makes the characterization of the WPFE more precise.

\vspace{3pt}
\textit{4) Privacy-preserving}
\vspace{3pt}

The distributed incremental update scheme is privacy-preserving, which is reflected in three aspects.

First, each MP does not need to collect the raw wind power and forecast data of other MPs. On the contrary, each MP only needs partly information from its neighboring MPs, i.e., $G_i^{t}$ in \eqref{ACA 1} and $\boldsymbol{Q}_i^{t}$ in \eqref{ACA 2}. Using these partly information as well as the consensus effect of the average consensus algorithm, each MP can eventually take the correlation between all DWGs into account.

Second, the information exchanged among adjacent MPs are privacy-preserving. No MP can deduce the raw data of other MPs from the exchanged information. Although in the 1st iteration of the average consensus algorithm, the $m$-th MP directly share its local calculation results to its neighbors, e.g., $\varrho_{j,m,i}$ in (\ref{local alpha 1}). However, the neighbors of the $m$-th MP cannot deduce $y_m^0$ from $\varrho_{j,m,i}$, as they do not know $\mu_{j,y_m}^{new}$. Other information exchanges among MPs are in the same situation.

Third, each MP can only obtain and update its own conditional PD using the distributed incremental update scheme. This is because that only the $m$-th MP knows and updates $\mu_{j,x_m}^{new}$. Thus, only the $m$-th MP can calculate the mean of its conditional PD, i.e., $\lambda_{j,m}^{new}$ in (\ref{new lambda}), by $\mu_{j,x_m}^{new}$. 

\vspace{10pt}
\noindent \textbf{6. Case Study} 
\vspace{5pt}

\begin{figure}[h]
\setlength{\abovecaptionskip}{0pt}  
  \centering 
  \includegraphics[width=3.3in]{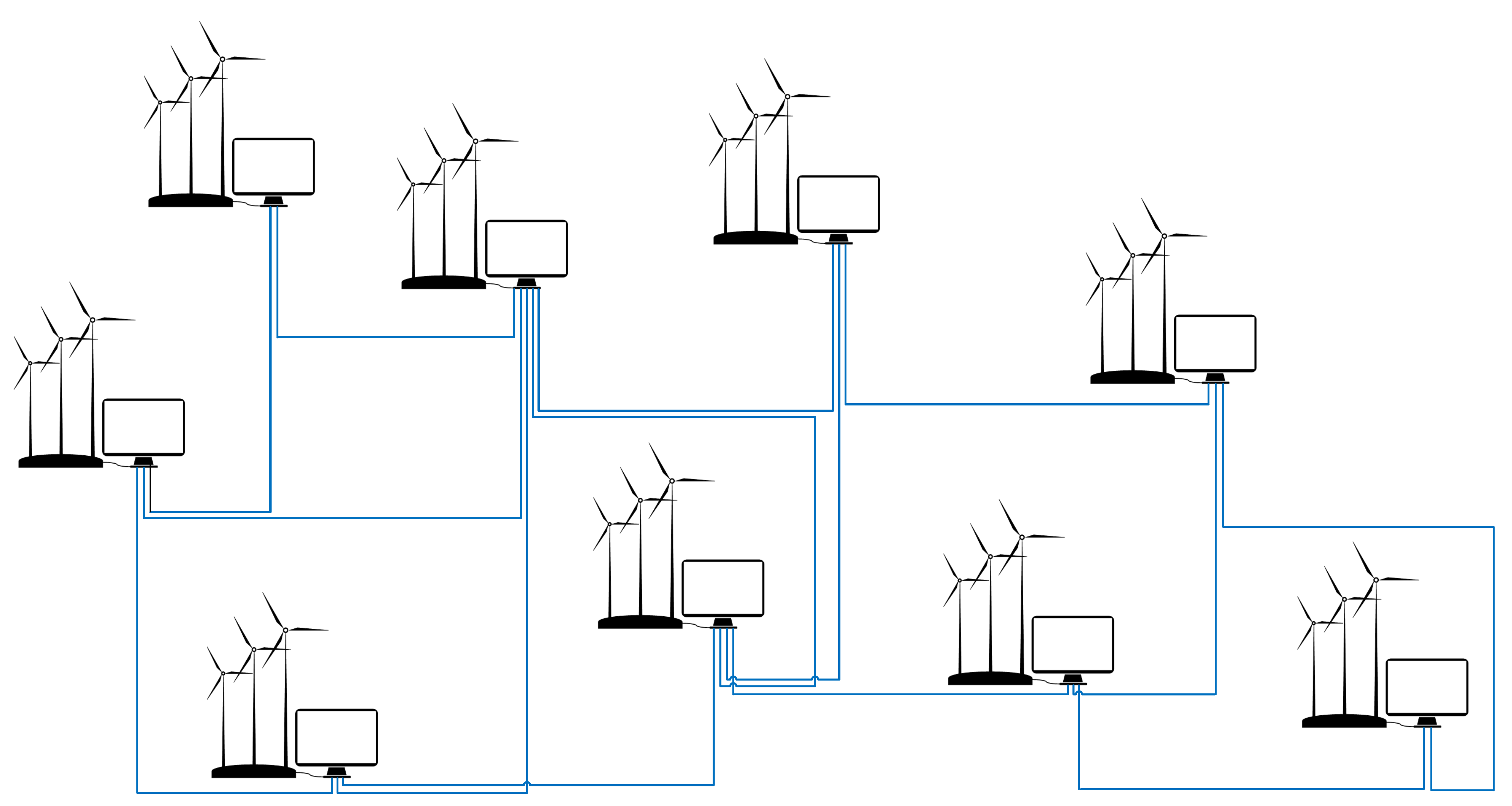} 
  \caption{Communication network topology for nine MPs}
  \label{case-topo1} 
\end{figure}

In this section, we first assume that there are nine MPs in a distribution network, and each of them has one DWG. The preset communication network topology of the nine MPs is shown in Fig.~\ref{case-topo1}, where each MP can only communicate with the MPs that directly connect to it. Then, we choose  40 days of hourly wind power and forecast data of nice sites from the Wind Dataset published by the National Renewable Energy Laboratory, the U.S.. These data are assumed as the historical dataset of the nine MPs. After that, we also choose the subsequent 40 days of hourly wind power and forecast data as the new dataset of for update. 

The main case studies are completed based on the above settings. Besides, the situations with more data and more DWGs are also investigated. 

\vspace{5pt}
\noindent \textbf{6.1 Verification of the Distributed Modified IGMM} 
\vspace{5pt}

For the chosen nine MPs, the distributed modified IGMM generates nine groups of updated results. We choose the 1st group, i.e., the updated result of the 1st MP, for illustration. Thus, we must first verify the rationality of this choice.

\begin{figure}[h]
\setlength{\abovecaptionskip}{0pt}
  \centering 
  \includegraphics[width=3.5in]{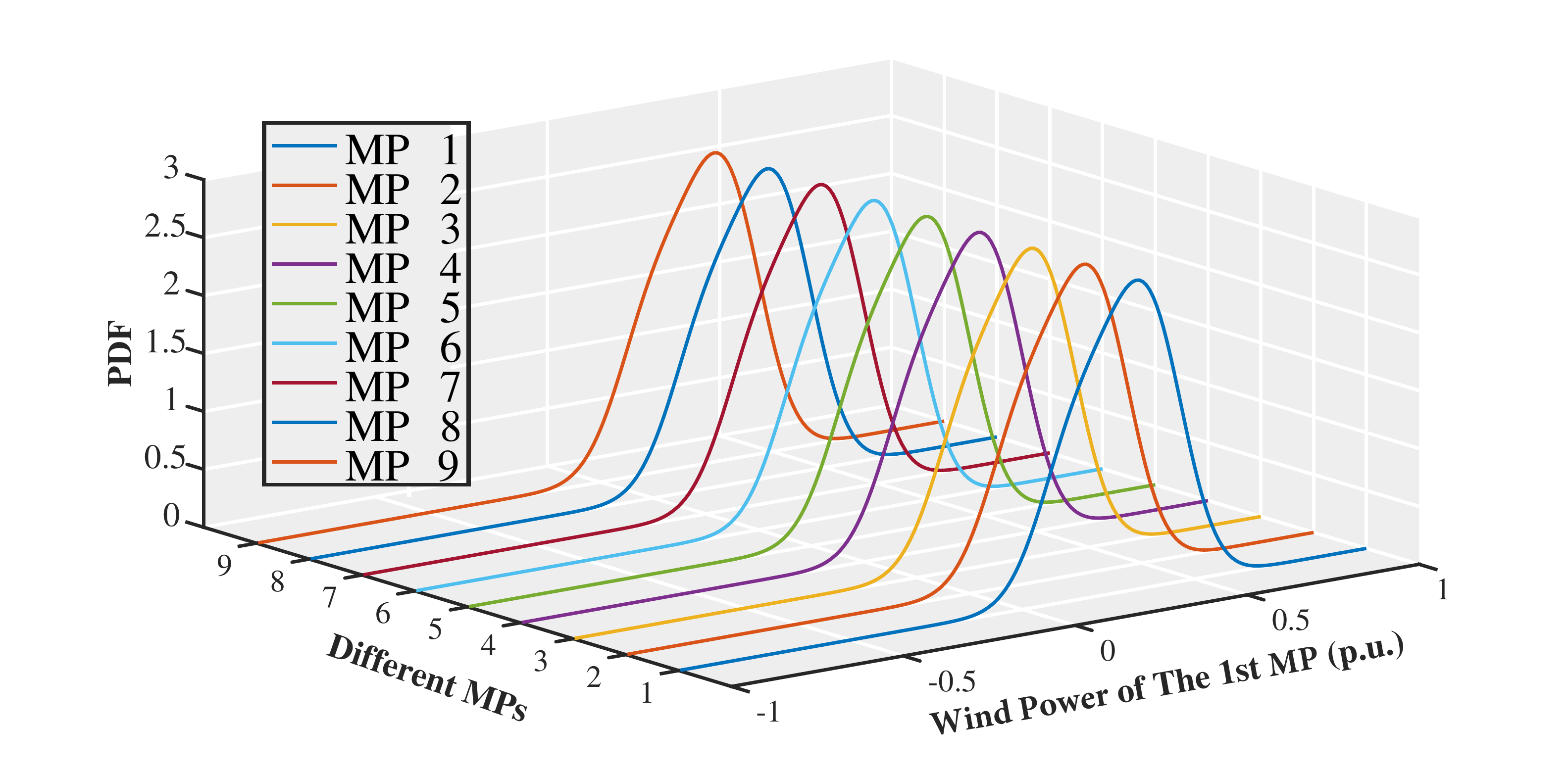} 
  \caption{The marginal PDF of the first dimension built by different MPs}
  \label{case-consensus} 
\end{figure}

Note that each MP does not own the complete updated parameter set of the joint PD, i.e., $\boldsymbol{\mu}_{j}$, $w_j$ and $\boldsymbol{\Sigma}_j$ for $j=1,...,J$, but only owns the updated $\mu_{j,x_m}$, $\mu_{j,y_m}$, $w_j$ and $\boldsymbol{\Sigma}_j$ for $j=1,...,J$. 
To illustrate the consensus effect of the distributed modified IGMM, we collect each MP's updated private mean value $\mu_{j,x_m}$ and $\mu_{j,y_m}$ to form a complete $\boldsymbol{\mu}_{j}$. Combining this $\boldsymbol{\mu}_{j}$ with each MP's $w_j$ and $\boldsymbol{\Sigma}_j$, we obtain nine complete updated parameter sets of the joint PD, which are essentially nine updated joint PDs. Then, we derive the marginal PD of the 1st dimension from each updated joint PD to illustrate the differences among the updated results of the MPs. The 9 marginal PDs are shown in Fig. \ref{case-consensus} in the form of the PD function (PDF). As shown, the PDFs built by different MPs are essentially the same. To quantify the differences, we calculate the Jensen--Shannon divergence between the complete updated joint PD built by the 1st MP and those by the other MPs and provide the results in Table \ref{Table_JS}. All the Jensen--Shannon divergences are lower than $9.83\times 10^{-12}$, which is an extremely low value relative to 1. Thus, the differences between the updated results of each MP are indeed small. Therefore, the consensus effect of the distributed modified IGMM is verified, and using the results of the 1st MP as a representative is also reasonable.

\begin{table}[h]
\setlength{\abovecaptionskip}{0pt}
	\renewcommand{\arraystretch}{1.3}
	\caption{The Jensen--Shannon Divergence Between the 1st MP and Other MPs}
	\label{Table_JS}
	\centering
	\footnotesize
	\setlength{\tabcolsep}{1mm}{
	\begin{tabular}{c c c c c c c c c c}
 	\toprule
	\bfseries MP & $1$ &  $2$ &  $3$ & $4$ & $5$ & $6$ &  $7$ &  $8$ & $9$ \\
	\bfseries Divergence ($\times 10^{-12}$) & 0 & 1.28 & 9.83 & 2.00 & 4.76 & 5.02 & 5.03 & 3.80 & 6.37 \\
	\hline
	\toprule
	\end{tabular}}
\end{table}

\begin{figure}[h]
\setlength{\abovecaptionskip}{0pt}  
  \centering 
  \includegraphics[width=3.4in]{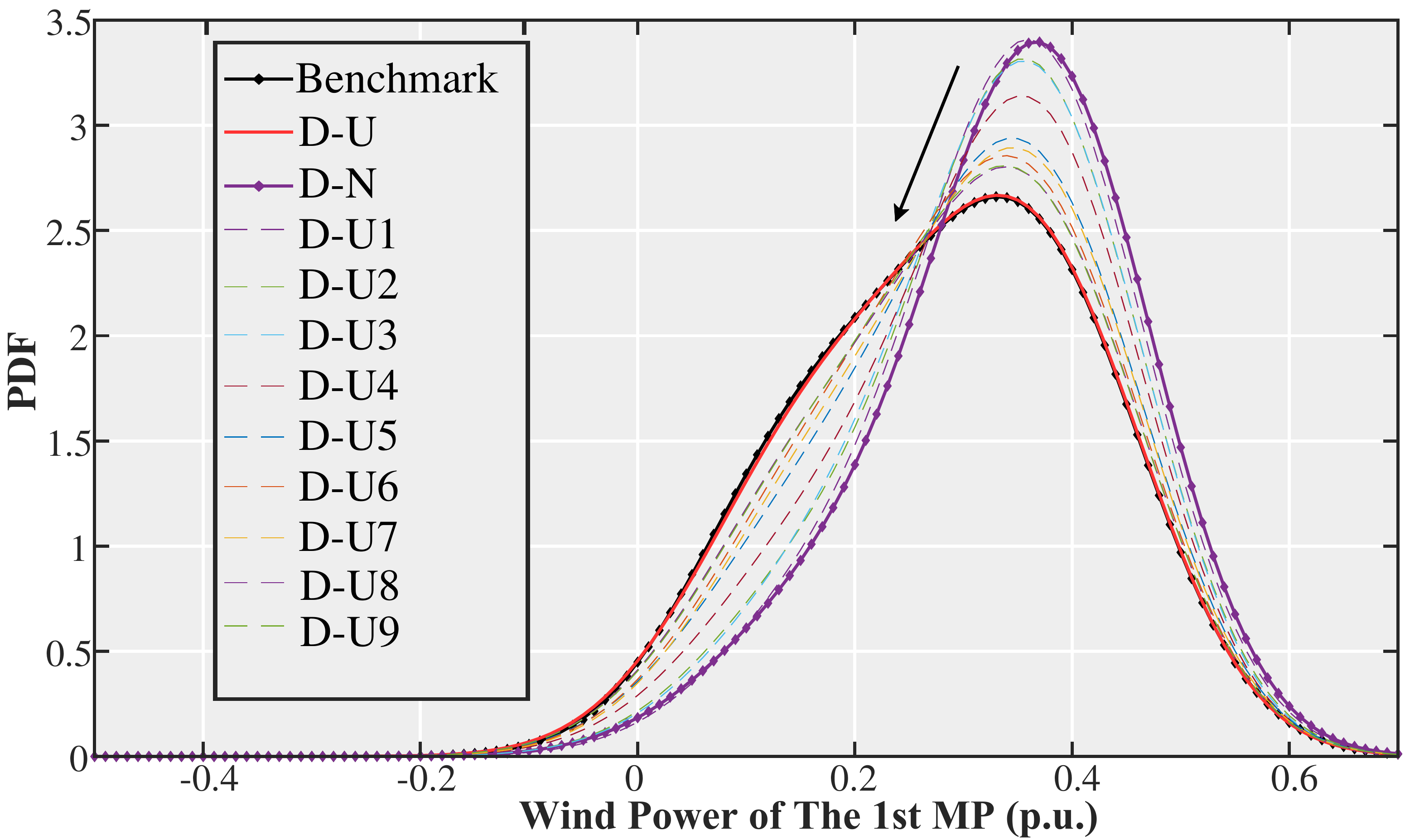} 
  \caption{The marginal PDF of the first dimension}
  \label{case update} 
\end{figure}

To verify the effectiveness and correctness of the distributed modified IGMM, we utilize the entire dataset, i.e., the combined historical and new dataset, as the training set, and we use the centralized EM algorithm for training. This training result is considered the final benchmark for the incremental update. Then, we use the proposed distributed modified IGMM to update the old PD, which is built using only the historical dataset. Because the new dataset has 960 pieces of new data, the distributed modified IGMM needs to be performed 960 times for the final complete update. For a clear illustration, we derive the marginal PD of the 1st dimension from the updated results. The marginal PDFs built by the centralized EM algorithm and distributed modified IGMM are shown in Fig. \ref{case update}. The benchmark is the result obtained by the centralized EM algorithm with the entire dataset. `D-U' represents the result of the distributed modified IGMM after 960 updates, and `D-N' represents the result of the old PD built by the historical dataset. The legend `D-U$n$' denotes the update result of the distributed modified IGMM after $n\times 96$ updates. From this illustration, we can draw two conclusions: (1) the effectiveness of the distributed modified IGMM is verified because as the number of updates increases, the updated PDF curves built by the distributed modified IGMM are gradually moving away from the `D-N' curve and approaching the benchmark in the direction of the black arrow. (2) The correctness of the distributed modified IGMM is verified because the curve `D-U' and the benchmark are coincident. 

\begin{figure}[h]
  \centering 
  \includegraphics[width=3.7in]{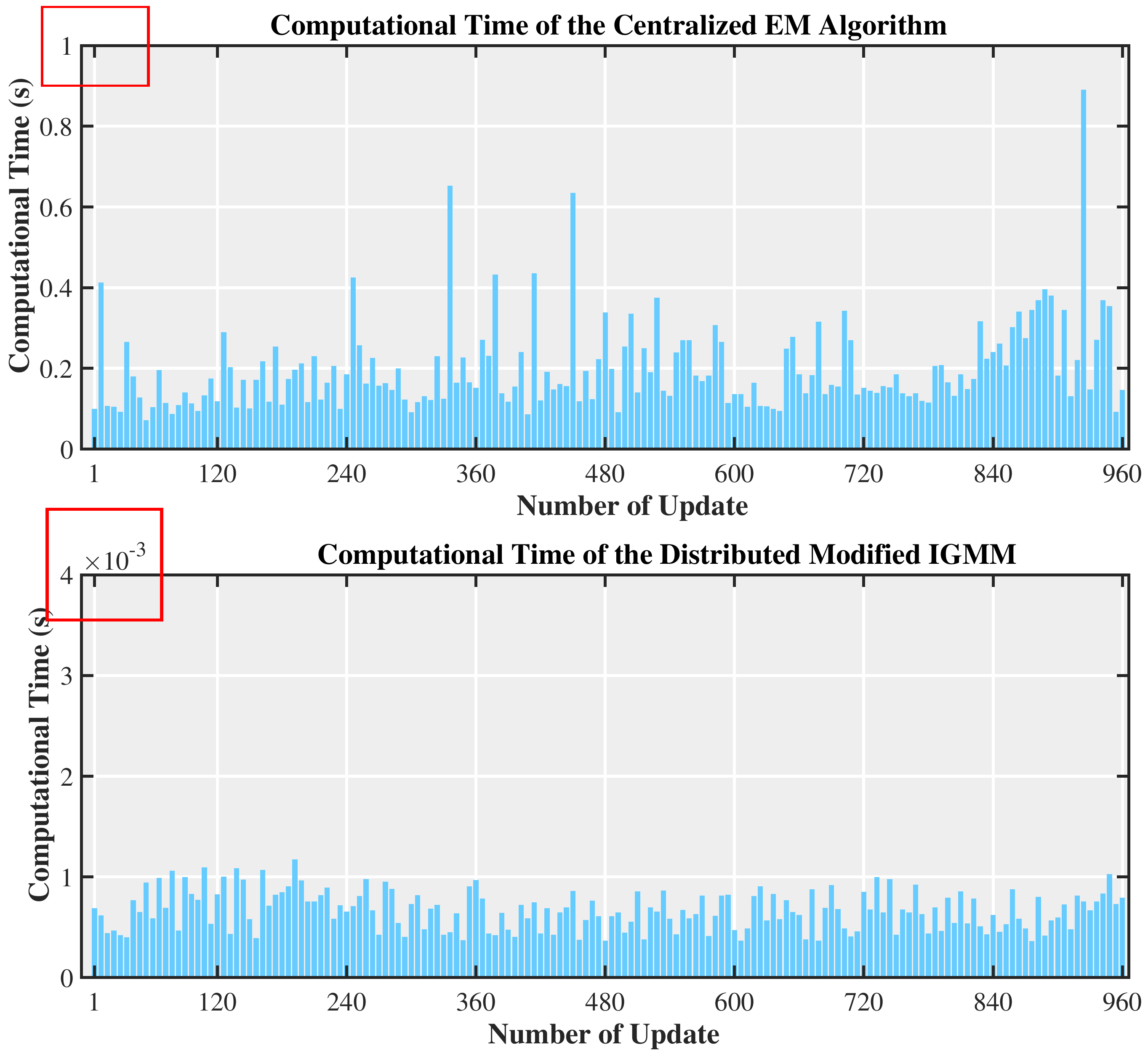} 
  \caption{The computational time for each update}
  \label{case-UpdateTime} 
\end{figure}

To verify the efficiency of the distributed modified IGMM, we compare the computational times of the centralized EM algorithm and the proposed distributed modified IGMM for each update. Note that the centralized EM algorithm has to reconstruct the training dataset whenever new data are generated and train this set to obtain the updated PD result. Besides, the calculation time of the proposed distributed modified IGMM refers to the time consumed by one MP, while the calculation time of the centralized EM algorithm refers to the time consumed by a data center. Although there are 960 updates, for a clear demonstration, we only illustrate the computational time for the 6th,  12th,  18th, ......, and 960th updates. The details are presented in Fig. \ref{case-UpdateTime}, which shows that the computational time of the distributed modified IGMM is much less than the computational time of the centralized EM algorithm. Meanwhile, as the size of the historical dataset increases, the computational time of the centralized EM algorithm also increases, but that of the proposed distributed modified IGMM is stable because the calculation burden of a piece of new data is always the same. Further comparisons of the computational time for all 960 updates are given in Table \ref{Computing Time comparison}. The results show that the proposed distributed modified IGMM is considerably more efficient than the centralized EM algorithm. 

\begin{table}[h]
\setlength{\abovecaptionskip}{0pt}
	\renewcommand{\arraystretch}{1.3}
	\caption{Computational Time comparison}
	\label{Computing Time comparison}
	\centering
	\footnotesize
	\begin{tabular}{c c c c}
	\hline
	\bfseries \ &  Centralized EM&  distributed modified IGMM &  Reduction\\
	\hline
	 Maximum & $1344\times 10^{-3}$s  & $1.22\times 10^{-3}$s  & 11006\% \\
	 Minimum & $69\times 10^{-3}$s  & $0.33\times 10^{-3}$s  & 20809\% \\
	 Mean    & $195\times 10^{-3}$s  & $0.67\times 10^{-3}$s  & 29004\% \\
	\hline
	\end{tabular}
\end{table}

\begin{figure}[h]
  \centering 
  \includegraphics[width=3.55in]{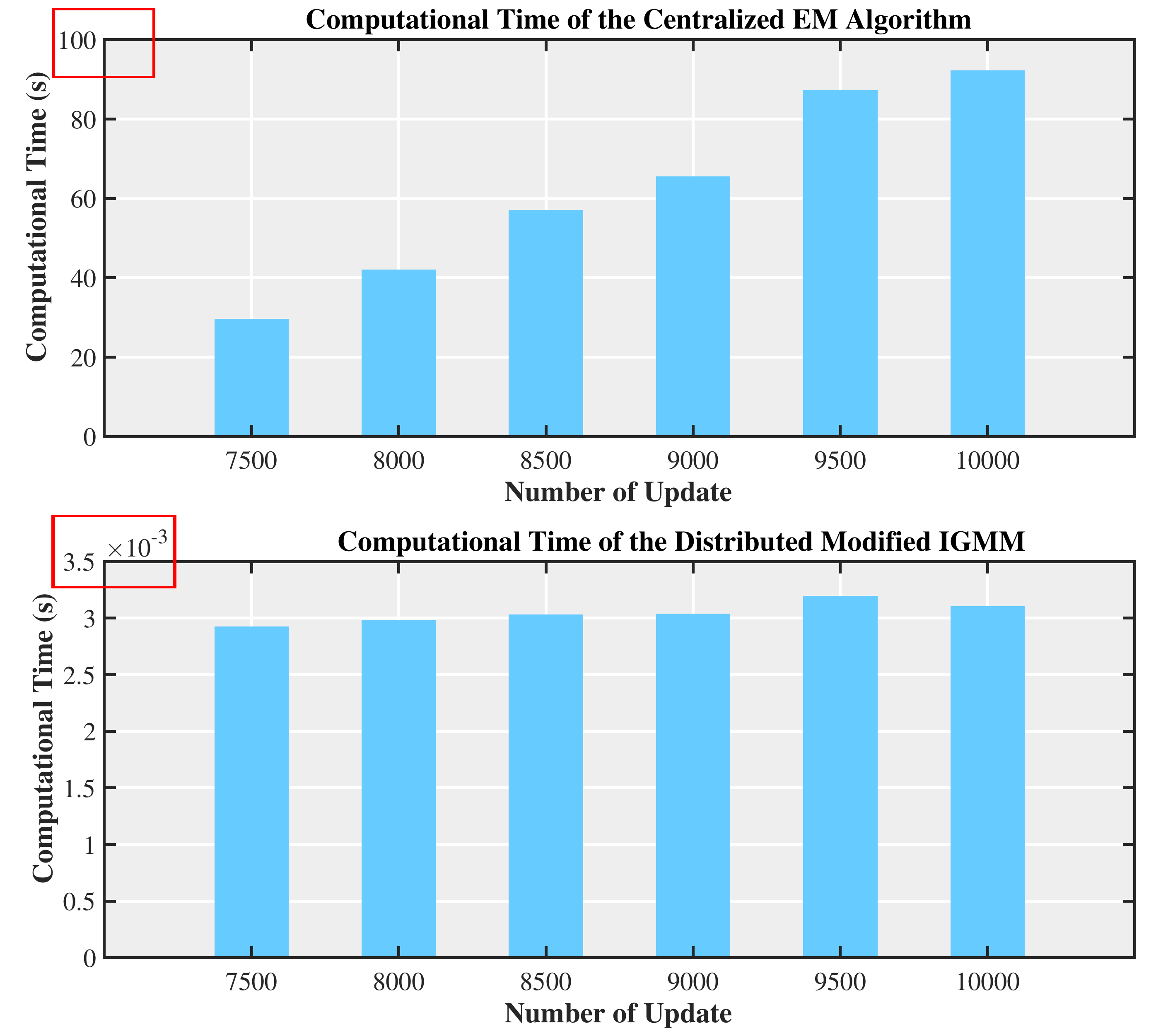} 
  \caption{The computational time for each update using a large dataset}
  \label{large computing time} 
\end{figure}

Note that even though the computational times of the EM algorithm are much higher than those of the proposed algorithm, the computational times are still lower than 1 second. This is because the dataset for testing consists of only 1920 pieces of observations. When the dataset grows over time with more integration of DWGs, the computational cost of the centralized EM algorithm will continue to increase quickly, but the computational cost of the proposed algorithm will still be much less than that of the EM algorithm and stable. To make a further comparison using a large training dataset, we choose 27 sites in Ohio from the Wind Dataset published by the National Renewable Energy Laboratory. The old dataset consists of 15,000 pieces of observation, while the new dataset consists of 10,000 pieces of observation. The computational times for the 7,500-th, 8,000th, ..., 10,000th updates are shown in Fig. \ref{large computing time}. 

From this illustration, we can draw these conclusions:
\begin{itemize}
	\item When a dataset grows, the computational time of the centralized EM algorithm increases sharply, e.g., the computational time for the 10,000th update approaches 100 seconds. Therefore, using the centralized EM algorithm for updating may indeed be a problem. 
	\item When the number of DWGs increases from 9 to 27, the computational time of the proposed algorithm increases slightly, i.e., from $1 \times 10 ^{-3}s$ to $3 \times 10 ^{-3}s$. Compared to the computational time of the 10,000th update using the centralized EM algorithm, the efficiency is improved by $2.9709 \times 10^{6} \%$. Therefore, the improvement made by the proposed algorithm is scalable when the dimension of the update problem increases.
	\item When the number of updates increases from 7,500 to 10,000, the computational times of the proposed algorithm remain low and stable. Compared to the centralized EM algorithm, the improvements vary from $1.0097 \times 10^{6} \%$ to $2.9708 \times 10^{6} \%$. Therefore, the improvement made by the proposed algorithm is scalable when the number of updates increases.
\end{itemize}

It should be emphasized that, due to the tiny computational cost of the distributed modified IGMM, this algorithm is sufficient for the needs of real time update. 

\vspace{5pt}
\noindent \textbf{6.2 Verification of the Distributed Incremental Update Scheme} 
\vspace{5pt}

Note that, the following case studies still use the small dataset consisting of 960 old observations and 960 new observations owned by nine MPs.

\begin{figure}[h]
\setlength{\abovecaptionskip}{0pt}  
  \centering 
  \includegraphics[width=3.5in]{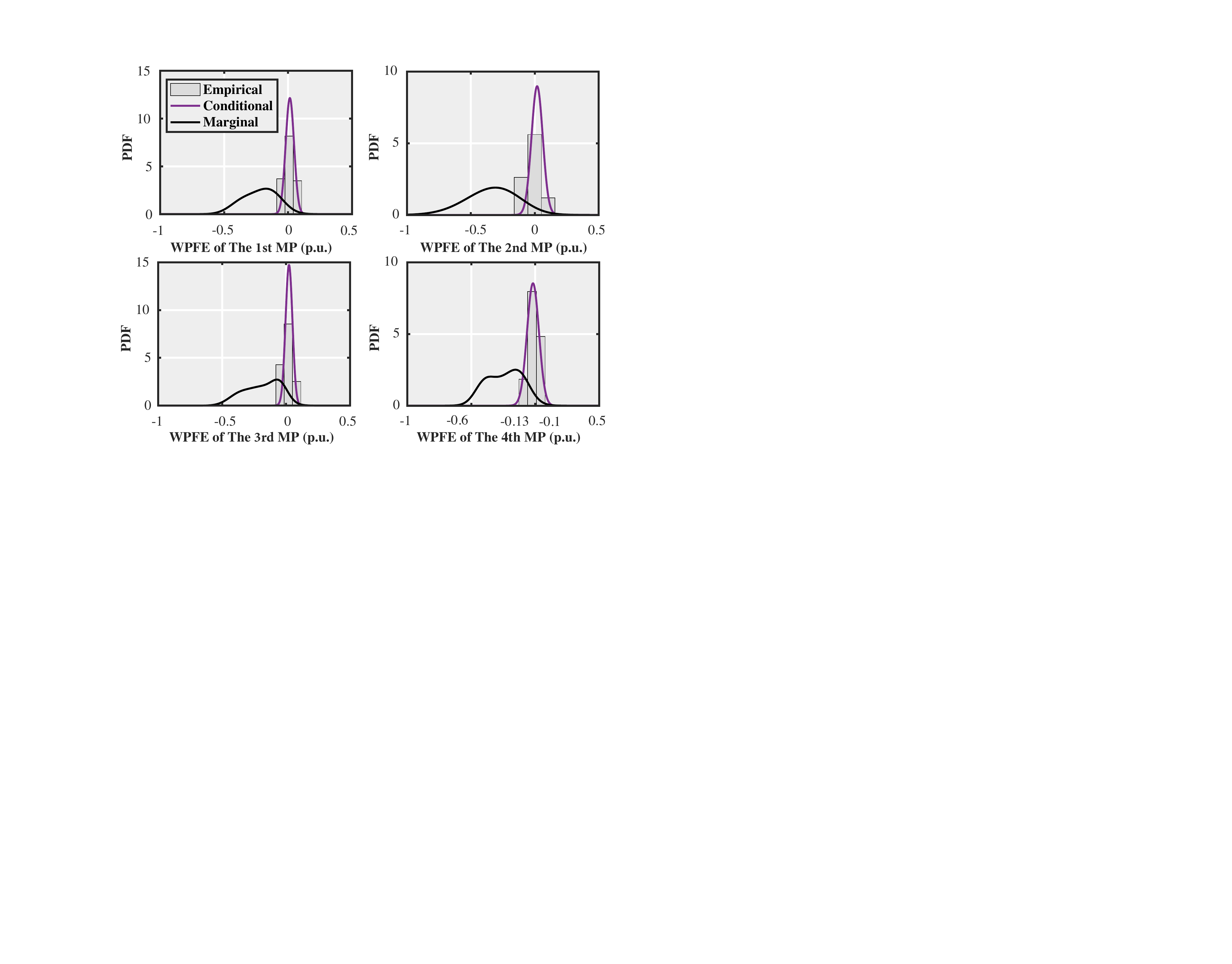} 
  \caption{comparison of PDFs}
  \label{case PCDF-MC} 
\end{figure}

To verify the effectiveness of the distributed incremental update scheme, we first use the entire dataset to build the empirical distributions under a given wind power forecast of correlated MPs. Then, we use the EM algorithm to build the latest marginal PD of each MP's WPFE based on the entire dataset without considering the correlation between DWGs. Finally, we use the proposed scheme to update each MP's conditional PD until the conditional PDs are the latest. The conditional PDs are shown in Fig. \ref{case PCDF-MC} in the form of PDFs. The conditional PDs of the first four MPs are shown. The updated conditional PDs built by the proposed scheme are more concentrated than the marginal PDs, making the characterization of the uncertainty more precise. This will greatly reduce the operation cost for each MP, e.g., the cost of reserve scheduling. Let's take the fourth subfigure in Fig. \ref{case PCDF-MC} for illustration. If the 4-th MP uses the updated marginal PDF to schedule reserves, a 0.6 (p.u.) down reserve and a 0.1 (p.u.) up reserve are required. However, if the conditional PDF is used, only a 0.13 (p.u.) down reserve and a 0.1 (p.u.) up reserve are required. Therefore, when using the updated conditional PDF for reserve scheduling, a huge cost of reserve will be saved. In addition, the updated conditional PDFs also match the empirical distributions well. 

\begin{figure}[h]
\setlength{\abovecaptionskip}{0pt}  
  \centering 
  \includegraphics[width=3.5in]{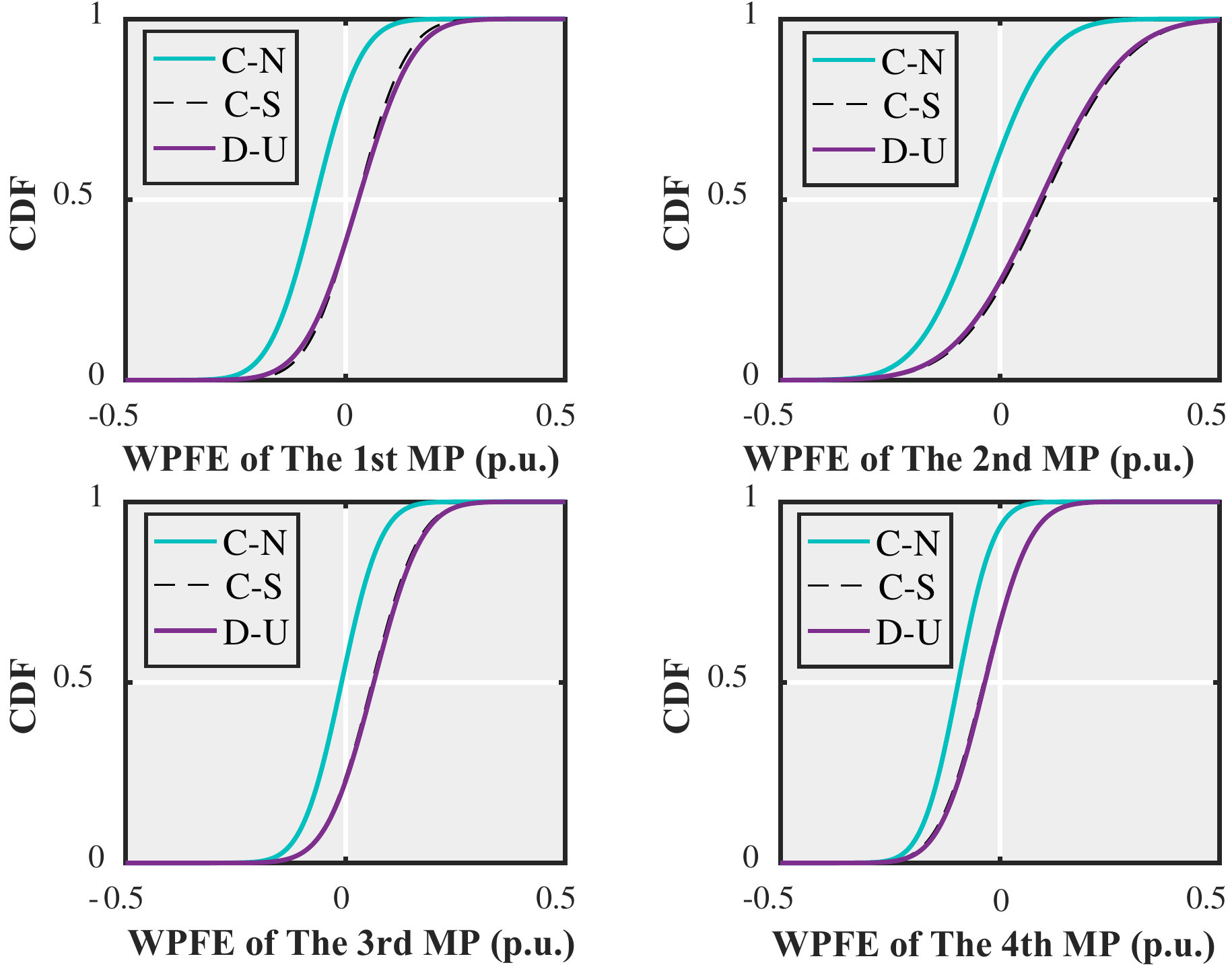} 
  \caption{comparison of CDFs }
  \label{case-4CDFs} 
\end{figure}

To verify the correctness of this scheme, we build each MP's conditional PD in different ways. First, we use a centralized EM algorithm to build the latest joint PD using the entire dataset and then derive each MP's conditional PD by collecting their wind power forecast data. These results are also considered as the benchmark and represented by the legend `C-S'. Second, we use the centralized EM algorithm to build the old PD using the historical dataset and then to derive each MP's conditional PD in a centralized manner. These results are represented by the legend `C-N'. Finally, we use the proposed scheme to update each MP's conditional PD until the conditional PDs are the latest. We use the legend `D-U' to represent these results. The conditional PDs of the first four MPs built by the three approaches are demonstrated in Fig. \ref{case-4CDFs} in the form of CDF.

\begin{table}[h]
\setlength{\abovecaptionskip}{0pt}
	\renewcommand{\arraystretch}{1.3}
	\caption{RSE comparison}
	\label{RSE comparison}
	\centering 
	\footnotesize
	\begin{tabular}{c c c c c}
	\hline 
	\bfseries  \ &   1st MP &   2nd MP &  3rd MP &  4th MP\\
	\hline 
	 C-N  & $6.68\%$  & $9.74\%$  & $3.78\%$  & $2.83\%$ \\
	 D-U  & $0.040\%$  & $0.038\%$  & $0.013\%$  & $0.009\%$ \\
	\hline 
	\end{tabular}
\end{table} 

As shown, curve `D-U' is coincident with the benchmark curve `C-S', while curve `C-N' differs greatly from the benchmark. The difference between the benchmark with curve `C-N' and the difference between the benchmark and curve `D-U' are measured by the relative standard error (RSE) and provided in Table \ref{RSE comparison}. Based on these illustrations, we know that the conditional PD updated by the scheme is correct because the RSE value between curve `D-U' and the benchmark is lower than $4\times 10^{-4}$, and curve `D-U' fits the benchmark.

\vspace{10pt}
\noindent \textbf{7. Conclusion} 
\vspace{5pt}

In this paper, we focus on MPs who own DWGs in a distribution network. To make them perform better in the local market considering the wind power uncertainty, we propose a distributed incremental update scheme to update each MP's conditional PD of the WPFE in a fully distributed manner with the consideration of the DWGs' correlation. This scheme consists of two original algorithms. One is a distributed modified IGMM for updating the parameters of the joint PD. Another one is a distributed derivation algorithm for deriving conditional PDs. This scheme can make each MP's conditional PD of the WPFE stay up to date with an extremely low and stable update cost, regardless of how much the historical dataset increases. In addition, this scheme is fully distributed without any data center or coordinator. Moreover, the conditional PD of the WPFE updated by the scheme is more concentrated and precise than the marginal PD, as the correlation among DWGs is taken into account. 

\vspace{10pt}
\noindent \textbf{Acknowledgements} 
\vspace{5pt}

This work is supported in part by the Joint Funds of the National Natural Science Foundation of China under Grant U1766206.

\vspace{20pt}



\bibliographystyle{IEEEtran}
\bibliography{IEEEabrv,paper}

\begin{thebibliography}{10}
\providecommand{\url}[1]{#1}
\csname url@samestyle\endcsname
\providecommand{\newblock}{\relax}
\providecommand{\bibinfo}[2]{#2}
\providecommand{\BIBentrySTDinterwordspacing}{\spaceskip=0pt\relax}
\providecommand{\BIBentryALTinterwordstretchfactor}{4}
\providecommand{\BIBentryALTinterwordspacing}{\spaceskip=\fontdimen2\font plus
\BIBentryALTinterwordstretchfactor\fontdimen3\font minus
  \fontdimen4\font\relax}
\providecommand{\BIBforeignlanguage}[2]{{%
\expandafter\ifx\csname l@#1\endcsname\relax
\typeout{** WARNING: IEEEtran.bst: No hyphenation pattern has been}%
\typeout{** loaded for the language `#1'. Using the pattern for}%
\typeout{** the default language instead.}%
\else
\language=\csname l@#1\endcsname
\fi
#2}}
\providecommand{\BIBdecl}{\relax}
\BIBdecl

\bibitem{morstyn2018using}
T.~Morstyn, N.~Farrell, S.~J. Darby, and M.~D. McCulloch, ``Using peer-to-peer
  energy-trading platforms to incentivize prosumers to form federated power
  plants,'' \emph{Nature Energy}, vol.~3, no.~2, pp. 94--101, 2018.

\bibitem{LUTH20181233}
\BIBentryALTinterwordspacing
A.~Lüth, J.~M. Zepter, P.~C. del Granado, and R.~Egging, ``Local electricity
  market designs for peer-to-peer trading: The role of battery flexibility,''
  \emph{Appl. Energy.}, vol. 229, pp. 1233 -- 1243, 2018. [Online]. Available:
  \url{http://www.sciencedirect.com/science/article/pii/S0306261918311590}
\BIBentrySTDinterwordspacing

\bibitem{LIU2018689}
\BIBentryALTinterwordspacing
Y.~Liu, K.~Zuo, X.~A. Liu, J.~Liu, and J.~M. Kennedy, ``Dynamic pricing for
  decentralized energy trading in micro-grids,'' \emph{Appl. Energy.}, vol.
  228, pp. 689 -- 699, 2018. [Online]. Available:
  \url{http://www.sciencedirect.com/science/article/pii/S0306261918309930}
\BIBentrySTDinterwordspacing

\bibitem{7930435}
H.~{Yi}, M.~H. {Hajiesmaili}, Y.~{Zhang}, M.~{Chen}, and X.~{Lin}, ``Impact of
  the uncertainty of distributed renewable generation on deregulated
  electricity supply chain,'' \emph{IEEE Trans. Smart Grid.}, vol.~9, no.~6,
  pp. 6183--6193, Nov 2018.

\bibitem{6481495}
Z.~{Zhang}, Y.~{Sun}, D.~W. {Gao}, J.~{Lin}, and L.~{Cheng}, ``A versatile
  probability distribution model for wind power forecast errors and its
  application in economic dispatch,'' \emph{IEEE Trans. Power Syst.}, vol.~28,
  no.~3, pp. 3114--3125, Aug 2013.

\bibitem{WU2014100}
\BIBentryALTinterwordspacing
J.~Wu, B.~Zhang, H.~Li, Z.~Li, Y.~Chen, and X.~Miao, ``Statistical distribution
  for wind power forecast error and its application to determine optimal size
  of energy storage system,'' \emph{INT J ELEC POWER.}, vol.~55, pp. 100 --
  107, 2014. [Online]. Available:
  \url{http://www.sciencedirect.com/science/article/pii/S0142061513003761}
\BIBentrySTDinterwordspacing

\bibitem{1490597}
A.~{Fabbri}, T.~G.~S. {Roman}, J.~R. {Abbad}, and V.~H.~M. {Quezada},
  ``Assessment of the cost associated with wind generation prediction errors in
  a liberalized electricity market,'' \emph{IEEE Trans. Power Syst.}, vol.~20,
  no.~3, pp. 1440--1446, Aug 2005.

\bibitem{EXIZIDIS201665}
\BIBentryALTinterwordspacing
L.~Exizidis, S.~J. Kazempour, P.~Pinson, Z.~de~Greve, and F.~Vallée, ``Sharing
  wind power forecasts in electricity markets: A numerical analysis,''
  \emph{Appl. Energy.}, vol. 176, pp. 65 -- 73, 2016. [Online]. Available:
  \url{http://www.sciencedirect.com/science/article/pii/S0306261916306468}
\BIBentrySTDinterwordspacing

\bibitem{7434078}
Z.~W. {Wang}, C.~{Shen}, and F.~{Liu}, ``Probabilistic analysis of small signal
  stability for power systems with high penetration of wind generation,''
  \emph{IEEE Trans. Sustain. Energy.}, vol.~7, no.~3, pp. 1182--1193, July
  2016.

\bibitem{7862254}
Z.~{Wang}, C.~{Shen}, F.~{Liu}, X.~{Wu}, C.~{Liu}, and F.~{Gao},
  ``Chance-constrained economic dispatch with non-gaussian correlated wind
  power uncertainty,'' \emph{IEEE Trans. Power Syst.}, vol.~32, no.~6, pp.
  4880--4893, Nov 2017.

\bibitem{wang2018}
Z.~Wang, C.~Shen, and F.~Liu, ``A conditional model of wind power forecast
  errors and its application in scenario generation,'' \emph{Appl. Energy.},
  vol. 212, pp. 771--785, 2018.

\bibitem{4470561}
F.~{Bouffard} and F.~D. {Galiana}, ``Stochastic security for operations
  planning with significant wind power generation,'' \emph{IEEE Trans. Power
  Syst.}, vol.~23, no.~2, pp. 306--316, May 2008.

\bibitem{6039388}
B.~{Hodge} and M.~{Milligan}, ``Wind power forecasting error distributions over
  multiple timescales,'' in \emph{Proc. IEEE PES General Meeting}, July 2011,
  pp. 1--8.

\bibitem{7779540}
Y.~{Wu}, P.~{Su}, and J.~{Hong}, ``An overview of wind power probabilistic
  forecasts,'' in \emph{Proc. IEEE PES Asia-Pacific Power and Energy
  Engineering Conference (APPEEC)}, Oct 2016, pp. 429--433.

\bibitem{6822599}
K.~{Bruninx} and E.~{Delarue}, ``A statistical description of the error on wind
  power forecasts for probabilistic reserve sizing,'' \emph{IEEE Trans.
  Sustain. Energy.}, vol.~5, no.~3, pp. 995--1002, July 2014.

\bibitem{5765544}
S.~{Tewari}, C.~J. {Geyer}, and N.~{Mohan}, ``A statistical model for wind
  power forecast error and its application to the estimation of penalties in
  liberalized markets,'' \emph{IEEE Trans. Power Syst.}, vol.~26, no.~4, pp.
  2031--2039, Nov 2011.

\bibitem{6202398}
N.~{Menemenlis}, M.~{Huneault}, and A.~{Robitaille}, ``Computation of dynamic
  operating balancing reserve for wind power integration for the time-horizon
  1–48 hours,'' \emph{IEEE Trans. Sustain. Energy.}, vol.~3, no.~4, pp.
  692--702, Oct 2012.

\bibitem{6074302}
W.~Lin, J.~Wen, S.~Cheng, and W.-J. Lee, ``An investigation on the active power
  variations of wind farms,'' in \emph{Proc. IEEE Industry Applications Society
  Annual Meeting}, Oct 2011, pp. 1--8.

\bibitem{6344672}
Z.~{Zhang}, Y.~{Sun}, J.~{Lin}, L.~{Cheng}, and G.~{Li}, ``Versatile
  distribution of wind power output for a given forecast value,'' in
  \emph{Proc. IEEE PES General Meeting}, July 2012, pp. 1--7.

\bibitem{7935491}
C.~{Tang}, J.~{Xu}, Y.~{Sun}, J.~{Liu}, X.~{Li}, D.~{Ke}, J.~{Yang}, and
  X.~{Peng}, ``A versatile mixture distribution and its application in economic
  dispatch with multiple wind farms,'' \emph{IEEE Trans. Sustain. Energy.},
  vol.~8, no.~4, pp. 1747--1762, Oct 2017.

\bibitem{5298967}
R.~{Singh}, B.~C. {Pal}, and R.~A. {Jabr}, ``Statistical representation of
  distribution system loads using gaussian mixture model,'' \emph{IEEE Trans.
  Power Syst.}, vol.~25, no.~1, pp. 29--37, Feb 2010.

\bibitem{8283770}
Z.~{Wang}, C.~{Shen}, Y.~{Xu}, F.~{Liu}, X.~{Wu}, and C.~{Liu}, ``Risk-limiting
  load restoration for resilience enhancement with intermittent energy
  resources,'' \emph{IEEE Trans. Smart Grid.}, pp. 1--1, 2018.

\bibitem{7744686}
M.~{Nijhuis}, M.~{Gibescu}, and S.~{Cobben}, ``Gaussian mixture based
  probabilistic load flow for lv-network planning,'' \emph{IEEE Trans. Power
  Syst.}, vol.~32, no.~4, pp. 2878--2886, July 2017.

\bibitem{8481551}
M.~{Jia}, C.~{Shen}, and Z.~{Wang}, ``A distributed probabilistic modeling
  algorithm for the aggregated power forecast error of multiple newly built
  wind farms,'' \emph{IEEE Trans. Sustain. Energy.}, pp. 1--1, 2018.

\bibitem{song2005highly}
M.~Song and H.~Wang, ``Highly efficient incremental estimation of gaussian
  mixture models for online data stream clustering,'' in \emph{Proc.
  Intelligent Computing: Theory and Applications III}, vol. 5803, 2005, pp.
  174--184.

\bibitem{declercq2008online}
A.~Declercq and J.~H. Piater, ``Online learning of gaussian mixture models-a
  two-level approach.'' in \emph{VISAPP (I)}, 2008, pp. 605--611.

\bibitem{chen2013incremental}
C.~Chen, D.~Mu, H.~Zhang, and W.~Hu, ``Incremental learning of gaussian mixture
  model: A top-down algorithm,'' \emph{Journal of Computational Information
  Systems}, vol.~9, no.~17, pp. 6971--6980, 2013.

\bibitem{bouchachia2011incremental}
A.~Bouchachia and C.~Vanaret, ``Incremental learning based on growing gaussian
  mixture models,'' in \emph{Proc. 10th IEEE International Conference on
  Machine Learning and Applications and Workshops}, vol.~2, 2011, pp. 47--52.

\bibitem{784637}
C.~{Stauffer} and W.~E.~L. {Grimson}, ``Adaptive background mixture models for
  real-time tracking,'' in \emph{Proc. 1999 IEEE Computer Society Conference on
  Computer Vision and Pattern Recognition (Cat. No PR00149)}, vol.~2, June
  1999, pp. 246--252 Vol. 2.

\bibitem{6491172}
A.~{Ribes}, J.~C. {Bueno}, Y.~{Demiris}, and R.~L. {de Mántaras},
  ``Context-gmm: Incremental learning of sparse priors for gaussian mixture
  regression,'' in \emph{Proc. IEEE International Conference on Robotics and
  Biomimetics (ROBIO)}, Dec 2012, pp. 1446--1451.

\bibitem{pinto2015fast}
R.~C. Pinto and P.~M. Engel, ``A fast incremental gaussian mixture model,''
  \emph{PloS one}, vol.~10, no.~10, 2015.

\bibitem{pinto2017scalable}
R.~Pinto and P.~Engel, ``Scalable and incremental learning of gaussian mixture
  models,'' \emph{arXiv preprint arXiv:1701.03940}, 2017.

\bibitem{jia2018privacy}
M.~Jia and C.~Shen, ``Privacy-preserving distributed joint probability modeling
  for spatial-correlated wind farms,'' \emph{arXiv preprint arXiv:1812.09247},
  2018.

\bibitem{4118472}
R.~{Olfati-Saber}, J.~A. {Fax}, and R.~M. {Murray}, ``Consensus and cooperation
  in networked multi-agent systems,'' \emph{Proc. of the IEEE}, vol.~95, no.~1,
  pp. 215--233, Jan 2007.

\end{thebibliography}

\end{document}